# Transformation numérique de l'éducation, approche systémique et recherche appliquée

## Digital Transformation of Education, Systems Approach and Applied Research

## Transformación digital de la educación, enfoque sistémico e investigación aplicada

Elie Allouche, chef de projet recherche appliquée
Ministère de l'Éducation nationale, France
elie.allouche@education.gouv.fr

RÉSUMÉ

Cet article propose la construction d'une modélisation systémique du numérique en éducation dans le cadre d'une recherche appliquée aux politiques publiques (ministère de l'Éducation nationale français). Considérant le numérique dans sa pervasivité, il met en évidence l'importance d'une approche complexe pour comprendre la transformation des pratiques. Comme modalité de recherche appliquée, nous présentons les groupes thématiques numériques (GTnum). L'approche méthodologique combine une posture réflexive éclairée par les apports de la recherche, des choix conceptuels centrés sur les humanités numériques et l'approche systémique, la recherche participative et la science ouverte via le carnet Hypothèses « Éducation, numérique et recherche ». Comme résultats, notre modélisation est centrée sur un « numérique environnant » et six unités d'action, mis à l'épreuve via les thématiques des GTnum. Nous interprétons ces résultats par une comparaison avec d'autres cadres systémiques, une application aux axes de la transformation numérique en académies, une réflexion prospective avec le développement de l'IA générative et des perspectives pour la recherche participative. Enfin, l'article discute des limites et apports de cette démarche : variabilité de la compréhension des enjeux et de l'intégration des apports de la recherche, mais pistes pour l'anticipation d'une nouvelle configuration du numérique avec la place de l'IA.

ABSTRACT

This article proposes the construction of a systemic model of digital education as part of research applied to public policy (French Ministry of Education). Considering the digital domain in its pervasiveness, it highlights the importance of a complex approach to understanding the transformation of practices. As an applied research modality, we present digital theme groups (GTnum). The methodological approach combines a reflexive posture informed by research contributions, conceptual choices centered on digital humanities and the systems approach, participatory research and open science via the Hypotheses "Education, digital and research" notebook. As a result, our modeling is centered on a "digital environment" and six units of action put to the test via the GTnum themes. We interpret these results through a comparison with other systemic frameworks, an application to the axes of digital transformation in academies, a prospective reflection with the development of generative AI and perspectives for participatory research. Finally, the article discusses the limits and contributions of this approach: variability in the understanding of the issues at stake and in the integration of research contributions, as well as avenues for anticipating a new digital configuration with the place of AI.

RESUMEN

Este artículo propone la construcción de un modelo sistémico de educación digital en el marco de una investigación aplicada a las políticas públicas (Ministerio de Educación francés). Teniendo en cuenta la omnipresencia de la tecnología digital, destaca la importancia de un enfoque complejo. Como método, presentamos los grupos temáticos digitales (GTnum). El enfoque metodológico combina una postura reflexiva informada por la investigación, las opciones conceptuales centradas en las humanidades digitales y el enfoque sistémico, la investigación participativa y la ciencia abierta a través del cuaderno de hipótesis "Educación, digital e investigación". Los resultados de nuestra modelización se centran en un "entorno digital" y en seis unidades de acción, puestas a prueba a través de los temas GTnum. Interpretamos estos resultados mediante una comparación con otros marcos sistémicos, una aplicación a los ejes de la transformación digital en las academias, una reflexión prospectiva con el desarrollo de la IA generativa y perspectivas para la investigación participativa. Por último, el artículo analiza los límites y las contribuciones de este enfoque: la existencia de variabilidad en la comprensión de los temas y en la integración de las contribuciones de la investigación, al mismo tiempo que pistas para anticipar una nueva configuración digital con el lugar de la IA.

# Introduction

L'objet de cet article est une recherche menée sur notre terrain professionnel, dans une posture de « praticien-chercheur » (De Lavergne, 2007) et dans un contexte de recherche appliquée[1] aux politiques publiques, avec comme principal objectif de comprendre la transformation numérique de l'éducation par une approche systémique. L'environnement professionnel est celui de nos missions à la DNE[2], qui consistent à accompagner la transformation numérique de l'éducation par la production de connaissances, la formation, l'accompagnement de projets associant chercheurs et praticiens de l'éducation, se traduisant par le dispositif fédérateur et partenarial des groupes thématiques numériques (GTnum) (MENJ, 2023a). Notre démarche réflexive s'inscrit dans une perspective empirico-conceptuelle (Lemieux, 2018) : dans sa dimension empirique, elle s'appuie sur une expérience professionnelle (pratiques du numérique dans l'enseignement, la formation et le management) et un point de vue institutionnel (ingénierie de projets et coordination de collectifs de recherche); dans sa dimension conceptuelle, étroitement liée à la mise à distance critique de l'objet, elle s'appuie sur – en même temps qu'elle interroge – les apports de la recherche sur le numérique en éducation dans ses dimensions sociale et épistémique.

Nous présenterons donc deux résultats : une contribution théorique avec une proposition de modélisation systémique du numérique tirant profit des apports des humanités numériques (HN) pour la réorganisation des savoirs dans une perspective interdisciplinaire ; une contribution méthodologique et pratique, en proposant d'évaluer la portée heuristique de cette modélisation comme cadre d'intelligibilité et d'action pour les politiques publiques.

Cette contribution a aussi une ambition prospective. En effet, notre modélisation se veut provisoire et évolutive, en visant non seulement l'analyse des transformations actuelles, mais aussi la compréhension de certaines tendances émergentes comme le développement des systèmes d'intelligence artificielle (IA) générative.

Nous suivrons donc les étapes suivantes : 1) une problématisation issue de notre itinéraire professionnel et de nos actions en recherche appliquée; 2) l'élaboration d'un cadre théorique basée sur une revue de littérature portant sur l'intelligibilité du numérique en éducation, pris comme un objet complexe; 3) la description des choix méthodologiques et du corpus mobilisé; 4) la présentation des résultats obtenus; 5) une discussion proposant une interprétation des résultats à quatre niveaux : comparatif, applicatif, méthodologique et prospectif.

---

[1] Par « recherche appliquée » nous entendons la « mise en œuvre pratique des connaissances [qui] exploite les avancées scientifiques et technologiques pour progresser dans un secteur d'activité donné » (Hcérès, 2020).

[2] Direction du numérique pour l'éducation (MENJ : ministère de l'Éducation nationale et de la Jeunesse). Pour le détail de ces missions, voir Allouche (2022).

# 1. Problématique

## 1.1 Contexte

Pour établir la problématique, nous partons de notre itinéraire professionnel, combinant cadre empirique et cadre conceptuel.

À partir de notre itinéraire professionnel et intellectuel, résumé en trois phases (dimension diachronique), la figure 1a permet d'identifier l'origine des données mobilisées et combinées pour déboucher sur une proposition de modélisation systémique. Ces données relèvent à la fois d'un cadre empirique (pratique professionnelle) et d'un cadre conceptuel (littérature scientifique) qui se nourrissent mutuellement (dimension synchronique).

**Figure 1a**

*Modélisation du numérique en éducation : cheminement parallèle du cadre empirico-conceptuel*

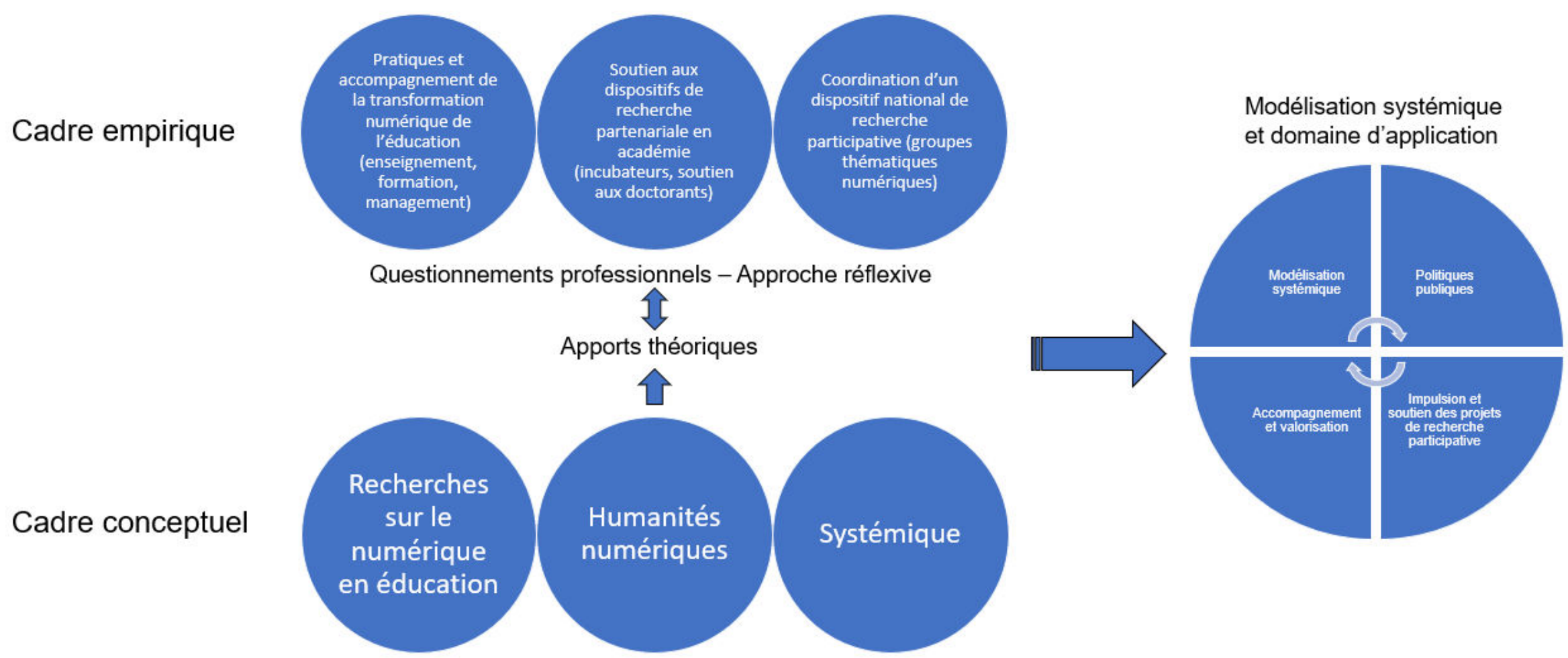

*Note*. © Elie Allouche.

**Figure 1b**

*Modélisation du numérique en éducation : processus de recherche appliquée*

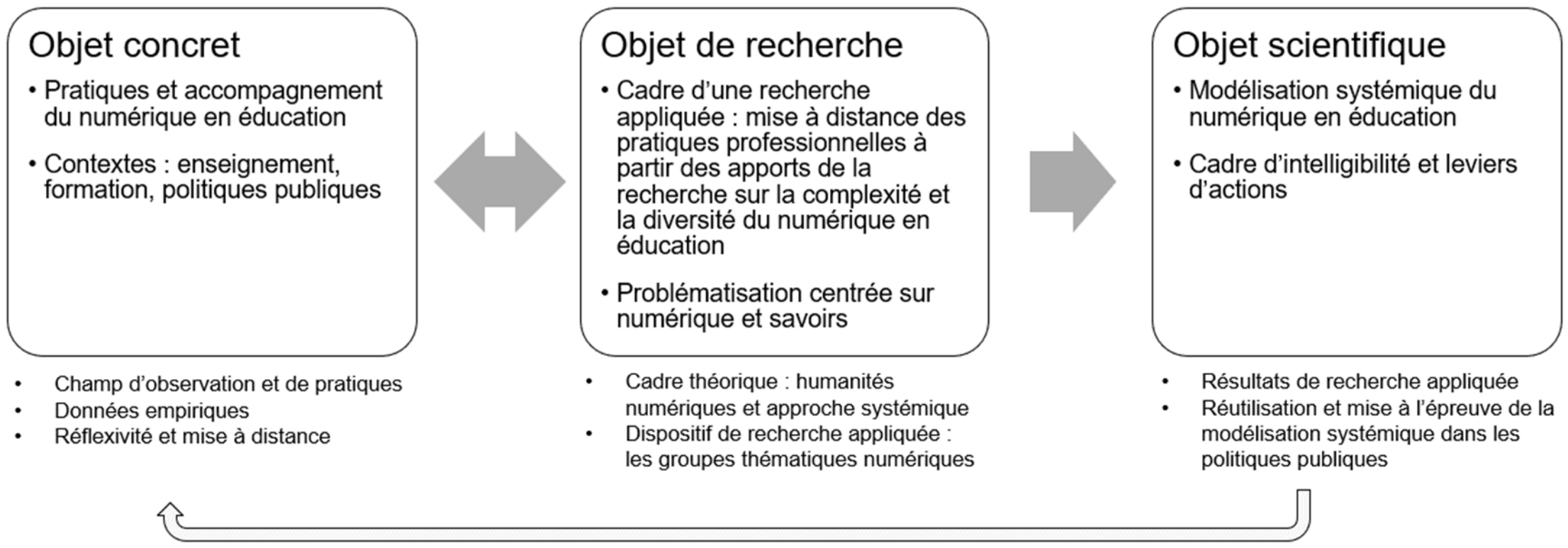

*Note*. © Elie Allouche inspiré de Davallon (2004).

La figure 1b relève aussi de cette double dimension en représentant plus précisément le processus de recherche appliquée. L'objet concret correspond au champ de la pratique professionnelle à partir duquel s'effectue la récolte de données empiriques. L'effort de réflexivité (prise de recul sur nos propres pratiques) et de mise à distance critique (travail d'objectivation) mène à l'identification de l'objet de recherche (complexité et problématisation) puis à l'objet scientifique en tant que tel, avec le travail de modélisation systémique visant un cadre d'intelligibilité et d'action. Une boucle de rétroaction, correspondant aux finalités de recherche appliquée telle que définie par Hcérès (2020), peut être alors établie avec la réutilisation de ce cadre dans le champ professionnel pour les politiques publiques.

Par ailleurs, cet itinéraire s'inscrit dans le cadre de la transformation numérique de l'éducation, qui peut être caractérisée par plusieurs composantes : contexte socioculturel, impulsion des politiques publiques avec le déploiement qui s'ensuit, et participation des acteurs eux-mêmes au processus de transformation (voir figure 2).

**Figure 2**

*Composantes de la transformation numérique de l'éducation*

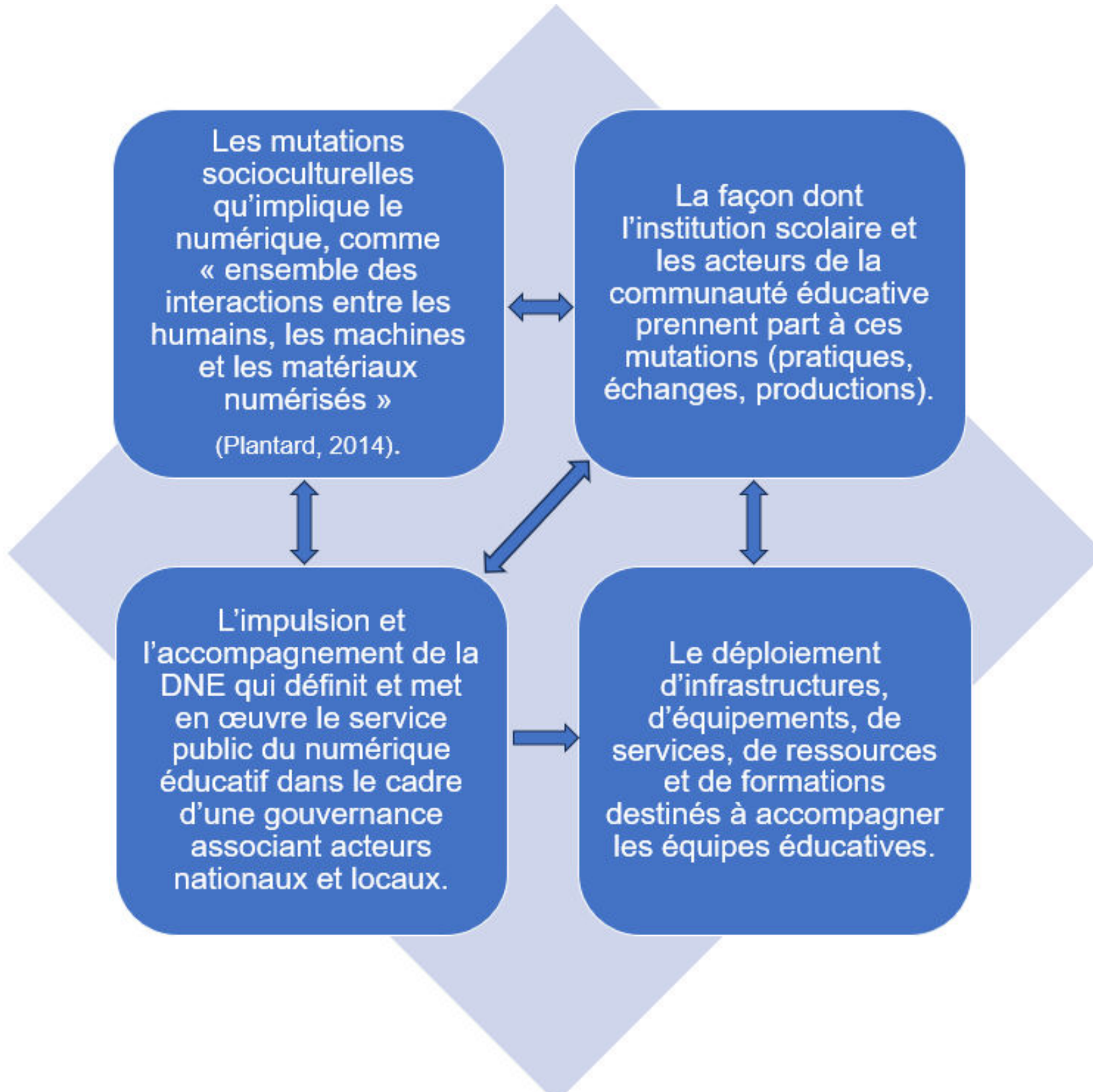

*Note*. © Elie Allouche adapté de Allouche (2022).

La problématique découle donc du caractère à la fois synchronique et diachronique de notre cheminement professionnel, interrogeant la capacité d'une recherche appliquée dans le domaine du numérique en éducation à produire des connaissances en vue d'une meilleure compréhension et intervention dans ce domaine.

## 1.2 Hypothèses

Afin de traiter la problématique, nous formulons ici trois hypothèses qui guideront la construction de notre cadre théorique puis nos choix conceptuels et méthodologiques.

- Une hypothèse théorique : à partir de son socle technologique, le numérique doit être considéré dans sa complexité et sa transversalité, chaque acteur ou groupe social focalisant son discours et ses actions sur une ou plusieurs dimensions qui lui semblent centrales en fonction des contextes3. Cette hypothèse intègre donc le constat d'une dispersion du traitement scientifique de ces aspects et la nécessité d'en proposer une synthèse articulée à l'hypothèse suivante.

- Une hypothèse théorique et pratique : construire une approche systémique et en proposer une modélisation doivent permettre de mieux comprendre et d'identifier les enjeux du numérique, actant le fait que nous serions passés d'une technologie à un environnement ou milieu qui façonne les pratiques individuelles et collectives, en étant façonné lui-même par ces pratiques. Cette approche peut être aussi une réponse à l'impasse de l'application telle quelle des recherches sur le numérique en éducation en raison de la complexité de l'objet qui appelle continuellement un effort de recontextualisation.

- Une hypothèse applicative : au titre de notre recherche appliquée, cette approche systémique peut venir en appui des politiques publiques en contribuant à l'intelligibilité de la transformation numérique de l'éducation comme objet complexe.

# 2. Cadre théorique

La première étape pour traiter notre problématique et vérifier ces hypothèses est la construction d'un cadre théorique pour identifier les éléments faisant du numérique en éducation un objet pluriel et complexe, répondant en cela à nos interrogations professionnelles, notamment sur la place et l'impact du numérique dans les domaines scientifiques, didactiques et pédagogiques (enseignement et formation), socioculturels et organisationnels (management et gestion de projet, actions partenariales).

## 2.1 Grille de lecture

Pour aborder l'état des connaissances, nous interrogeons la façon dont la littérature peut contribuer au besoin d'intelligibilité du numérique en éducation, impliquant de multiples acteurs et problématiques, sans essentialiser les technologies, mais en favorisant la contextualisation, l'objectivation des situations locales et l'agentivité des acteurs[4].

La construction de notre cadre théorique part donc des questions les plus générales, considérant le numérique en éducation dans sa genèse avant d'examiner les études qui élargissent son périmètre aux implications sociales, anthropologiques, culturelles et épistémologiques.

## 2.2 Mise en perspective historique

On peut se placer d'abord dans la perspective du temps long d'une histoire des « machines à enseigner » (Bruillard, 1997) et des « industries éducatives » (Moeglin, 2010). L'informatique à l'école en tant que telle débute dans les années 1970 (« expérience des 58 lycées ») et fait l'objet d'un rapport de l'INRP[5] rédigé par Baron *et al.* (1981), qui y voient le signe d'une « prise de conscience d'un phénomène de société » avec comme enjeu de « faire entrer une science et une technologie nouvelles dans la culture générale de l'homme du XX^e^ siècle[6] ». L'informatique est ensuite intégrée dans les TIC (technologies de l'information et de la communication, englobant audiovisuel, informatique et télématique) approximativement à partir de 1990, dont les « applications éducatives » constituaient un « objet de recherche et [un] champ de pratiques non stabilisé où aucune théorie n'a de position dominante » (Baron, 1994).

[3] Par exemple une collectivité au moment d'équiper un territoire scolaire, un chercheur au moment de concevoir un projet de recherche ou un enseignant dans le traitement didactique et l'instrumentation d'une séquence pédagogique pour ses élèves.

[4] Pour le détail de ceux-ci, on peut consulter la cartographie des acteurs établie par la DNE dans le cadre de sa stratégie (MENJ, 2023b) p. 10-11.

De même, Bruillard *et al.* (2002) partent de l'ensemble des technologies de l'information en y distinguant spécifiquement l'informatique comme « discipline scolaire, technologie éducative et ensemble d'instruments », tandis que Baron *et al.* (2018) mettent en avant « la diversité des approches sous-tendant les usages des [TIC] en éducation ».

## 2.3 Des TICE au numérique en éducation

En évaluant approximativement la fréquence des termes utilisés dans la littérature francophone sur la période 1970-2019, hors éducation, nous observons un basculement en faveur du terme « numérique » en 2009 (Ngram Viewer, 2023)[7] avec comme hypothèse la plus probable le fait d'y déceler le signe du développement d'un « numérique ambiant » et « pervasif » (Boullier, 2016 ; Delmas-Rigoutsos, 2018) dominé par les objets connectés et les réseaux sociaux à partir du début des années 2000.

Néanmoins, dans l'éducation comme dans les autres secteurs d'activité, nous constatons une acception très diverse, hétérogène, parfois contradictoire, du terme utilisé aussi bien comme adjectif que comme substantif, investi de multiples façons, comme synonyme d'outils ou comme un ensemble plus large.

La formulation « numérique éducatif » interroge également la recherche : Baron et Boulc'h (2011) ou Baron (2014), récemment repris par Fluckiger (2020) et CNESCO (2020), y voient essentiellement une émanation des pouvoirs publics pour désigner le champ vaste et pluriel de l'utilisation des technologiques informatiques dans l'éducation, dans lequel Inaudi (2017b) inclut les matériels et les contenus, en mobilisant « une diversité d'acteurs », tandis que pour Ailleri (2017) le choix du substantif est significatif d'un rapprochement entre matériels et transformation des pratiques[8].

Ces analyses révèlent donc un certain flottement terminologique et sémantique, en même temps qu'une pluralité d'objets d'étude à considérer. Par ailleurs des publications de portée internationale qui dressent régulièrement un état des lieux (comme UNESCO, 2023) témoignent d'une tension entre essentialisation de technologies considérées implicitement comme « auto-agissantes » et sommées de faire leurs preuves sur l'efficacité des apprentissages, et prise en compte de la diversité des acteurs et des contextes locaux ou nationaux, sociaux et institutionnels.

---

[5] Institut national de recherche pédagogique, devenu Institut français de l'éducation.

[6] Sur l'histoire de l'informatique dans l'enseignement en France, on peut s'appuyer sur les repères posés par Baron (2014), Pélisset (1985), et Inaudi (2017a), et sur le numérique dans les discours et textes officiels sur Rouissi (2017).

[7] Malgré les biais importants de l'outil Google Books Ngram, son utilisation dans le cas présent n'a d'autre ambition que de donner qu'un indicateur lexicométrique macro. Cet indicateur est confirmé par la base Google Scholar (au 01/09/23) en comparant, sur la période 2000-2023, la fréquence des termes « numérique école » (56700), « informatique école » (29000) et « TICE » [Technologies de l'Information et de la Communication pour l'Enseignement] (19900). Cette domination du terme « numérique » (comparé aux termes « TICE » et « informatique ») est confirmée par l'outil lexicométrique (Azoulay et De Courson, 2021) sur le corpus « Cairn.info » en sciences de l'éducation (50 % de fréquence dans le corpus en 2000, près de 70 % en 2020).

[8] Il est à noter ici que du côté des politiques publiques un arbitrage terminologique est effectué dans la stratégie de la DNE (MENJ, 2023b) : le « numérique éducatif » s'organise autour des axes suivants : gouvernance, animation, infrastructures, équipements, services, outils, ressources, formation aux compétences, tandis que le « numérique pour l'éducation » constitue l'ensemble global numérique éducatif + conduite du changement, innovation, activités de support.

Proposant un bilan des recherches, Fluckiger (2017) note en effet « l'impasse » que constituent « la mesure et/ou l'évaluation d'une efficacité éventuelle des technologies pour mieux apprendre » et l'absence de lien direct de causalité entre innovation technologique et innovation pédagogique. Manifestement les pratiques étudiées, dans leur diversité[9], ne se construisent pas sur un terrain vierge, mais s'intègrent dans un tissu social et relationnel, dans des pratiques personnelles et professionnelles préexistantes, et relèvent de choix et de stratégies d'acteurs contextualisés (Plantard, 2023).

## 2.4 Vers un élargissement des perspectives aux dimensions sociales et épistémiques du numérique

À partir de ce constat d'hétérogénéité des approches et des objets, faut-il alors parler d'effets « du » numérique ou du numérique comme réalité socioculturelle multiple et dispersée?

De fait, le panorama établi par Baron et Depover (2019) rend déjà compte d'une grande diversité d'objets d'étude et de problématiques qui relèvent à la fois de questions instrumentales, cognitives, pédagogiques, didactiques, institutionnelles ou sociales. Malgré une acceptation à dominante technologique, au moins dans la qualification des thématiques, plusieurs études soulignent les implications larges de ce champ, notamment pour la culture scolaire (Fluckiger, 2016) ou la forme scolaire (Cerisier, 2015), reprises en cela par un rapport de l'Inspection générale (Bechetti-Bizot, 2017). En réponse à une approche « techno-centrée », Ceci (2018) présente le numérique éducatif comme « l'association d'un outil, d'une culture et d'une pédagogie adaptée ».

Pour dépasser une approche déterministe, il s'agirait alors de passer à l'étude de l'« environnement numérique » (Hardouin *et al.*, 2018), du « dispositif » ou de « l'instrument », « permettant de penser le technique et le social dans leur relation dialectique » (Fluckiger, 2019).

En ce sens, pour Plantard (2014), reprenant Marcel Mauss, le numérique se définit d'abord comme un « fait social total » centré sur les interactions « entre les humains, les machines et les matériaux numérisés ».

Les repères théoriques que propose Collin (2023) rejoignent cet élargissement de perspectives sur le « façonnage social des technologies » : nature sociale des technologies, nature technique de la société, rôle des objets techniques dans les interactions sociales et la stabilisation de celles-ci; prise en compte du caractère non linéaire et ambivalent des processus sociotechniques, notamment d'innovation (choix, négociations, compromis ou contraintes en jeu, multiplicité d'acteurs aux intérêts divers dotés de pouvoir d'agir inégal).

[9] Dont (Fluckiger, 2020) donne un aperçu général sous le terme « usages du numérique » dans son propos introductif, reprenant notamment la méta typologie de (Basque et Lundgren-Cayrol, 2002).

Cette complexification du questionnement peut également s'appuyer sur l'apport de deux auteurs : Gilbert Simondon, pour qui « la présence de l'homme aux machines est une invention perpétuée », les machines intégrant « de la réalité humaine, du geste humain fixé et cristallisé en structures qui fonctionnent » (Simondon, 1989), et qui établit le concept de « milieu associé », environnement proche dans lequel se constitue l'individu (processus d'individuation[10]); Bertrand Gille, pour la caractérisation d'un « système technique » : 1) les techniques forment le système, en étant dépendantes les unes des autres, 2) elles sont liées aux autres « systèmes » constitutifs d'une société (social, économique, politique, juridique, etc.) (Gille, 1978 ; Lemonnier, 1983).

Pour nous centrer sur le domaine des savoirs, commun à l'éducation, la formation et la recherche, le constat de ces enjeux plus larges nous conduit à examiner les apports des HN dont le champ d'études porte sur le numérique à partir des pratiques en sciences humaines et sociales (SHS).

## 2.5 Apports des humanités numériques pour éclairer les enjeux du numérique en éducation

En effet, résumé à ses trois caractéristiques majeures dans la recherche[11] (voir tableau 1, colonne de gauche), le champ des HN peut éclairer à plusieurs titres les enjeux du numérique en éducation et pour la formation (colonne de droite).

---

[10] « L'individu n'est pas isolé sur lui-même, il porte plus que lui-même, une réserve d'individuation qui réside dans le couple qu'il forme avec son milieu associé. » (Roux, 2004).

[11] D'après ENS PSL (2021).

**Tableau 1**

*Les humanités numériques entre recherche, éducation et formation*

| Recherche | Éducation et formation |
|---|---|
| Une transdiscipline formée à partir des pratiques numériques en SHS, en s'inscrivant dans le temps long de l'écrit et de la production des savoirs. | Les pratiques inter et transdisciplinaires du numérique en éducation. |
| Un objet de recherche sur le tournant numérique de la société et ses implications pour la production et la diffusion des savoirs. | Cet objet de recherche inclut les enjeux éducatifs et formatifs, la production et la diffusion des savoirs étant des objets communs entre recherche et éducation. |
| Une communauté de praticiens, ouverte, pluridisciplinaire et internationale. | La dynamique de regroupement en communauté de praticiens concerne aussi les acteurs de l'éducation et de la formation. |

*Note*. © Elie Allouche.

Cette dynamique collective étant centrale, les HN sont donc définies non pas comme une entité figée, mais comme un processus collectif : un « dialogue interdisciplinaire sur la dimension numérique des recherches en sciences humaines et sociales, au niveau des outils, des méthodes, des objets d'études et des modes de communication » (Dacos et Mounier, 2014). Orientées vers la pérennisation de l'accès aux savoirs, elles se matérialisent notamment par des revues (Humanistica, 2020; Massot *et al.*, 2023), des infrastructures (Huma-Num, 2015), des publications de corpus et des plateformes favorisant l'accès ouvert (Cantrel, 2018; Carlin et Laborderie, 2021), des temps d'échange plus ou moins formalisés et des formations (Allouche et Desfriches-Doria, 2021; Dariah Clarin, 2014).

Dans leur structuration et leur institutionnalisation, les HN font ainsi écho aux problématiques rencontrées dans l'éducation, comme l'attestent plusieurs des articles du manifeste des HN (Dacos, 2010) qui rejoignent des finalités de l'éducation et de la formation (Allouche, 2020). Une esquisse de modélisation des formations en HN peut aussi aider à identifier les principaux thèmes abordés et les compétences visées, dans les contenus comme dans les dispositifs d'ingénierie pédagogique (Allouche et Desfriches-Doria, 2021). Parmi les thèmes communs de ces formations, figurent en effet le recueil, la production et le traitement des données, thème qui correspond au premier domaine du cadre de référence des compétences numériques (CRCN)[12].

Ainsi plusieurs publications émanant des SHS peuvent aider à mieux comprendre le numérique dans ses implications sociales, culturelles, documentaires ou épistémologiques (Bourdeloie, 2014).

Par ailleurs, à partir de sa notion d'« humanisme numérique », Doueihi (2013) définit le numérique comme le stade d'une évolution de l'informatique vers une « culture » (après avoir été successivement « branche des mathématiques », « nouvelle science à part entière » puis « industrie »), tandis que Vitali-Rosati (2014) identifie le numérique à « l'espace dans lequel nous vivons », correspondant à « un environnement dans lequel nous sommes plongés, qui détermine et façonne notre monde et notre culture ».

[12] https://eduscol.education.fr/721/evaluer-et-certifier-les-competences-numeriques

Reprenant les travaux de Simondon, Stiegler (2015) estime que « l'écriture et la lecture numériques constituent le nouveau milieu des savoirs », tandis que Merzeau (2013), Merzeau et Mulot (2017) voient dans le numérique « un milieu beaucoup plus qu'un outil », avec « l'introduction d'une dimension culturelle, sociale et éthique (…) dans une perspective non plus techniciste, mais historique et citoyenne ».

Enfin, Baron (2020) remet en perspective ces questionnements en rappelant quelques étapes du rapprochement entre SHS et informatique, y compris dans l'enseignement scolaire français avec le thème « informatique et société » dès les années 1980.

## 2.6 L'approche systémique

Prenant acte de cette complexité d'un objet technologique identifié aussi comme environnement social et culture, viser une modélisation systémique du numérique en éducation répond ainsi à un besoin d'intelligibilité pour les politiques publiques, celles-ci accompagnant la diversité des situations collectives et individuelles.

D'après Diemer (2014), on trouve parmi les fondements de l'approche systémique et de la pensée complexe :

- le « rapprochement de plusieurs disciplines dont la cybernétique » (Wiener, 1948), « la théorie de l'information » (Shannon et Weaver, 1949) et « la théorie des systèmes » (Von Bertalanffy, 1954);

- une méthodologie « permettant de rassembler et d'organiser les connaissances en vue d'une plus grande efficacité de l'action » (Rosnay, 1975);

- la notion de système (Morin, 2005), comme « ensemble d'éléments en interaction dynamique, organisés en fonction d'un but (…), outil de modélisation permettant de représenter et d'analyser des complexes d'éléments caractérisés par leur nombre élevé et un réseau de relations imbriquées » (Forrester, 1965);

- une « démarche globale qui met l'accent sur le relationnel plus que sur les objets, et qui permet d'appréhender la complexité d'un problème ».

Les principes du paradigme systémique sont en outre rappelés par Althaus *et al.* (2013) :

- causalité circulaire (« les effets agissent sur leurs propres causes et réciproquement »);

- principe téléologique (« mettre en avant le sens et la finalité de toute démarche intellectuelle »);

- globalisme (« considérer l'objet étudié comme une partie d'un tout, indissociable de son environnement »);

- agrégativité (« ne s'intéresser qu'à des éléments-clés [de la réalité], sélectionnés selon leur utilité et leur pertinence pratique »).

Appliqué au « système technicien », un système en sciences sociales est caractérisé par Ellul (1977) comme un « ensemble d'éléments en relation les uns avec les autres de telle façon que toute évolution de l'un provoque une évolution de l'ensemble ».

Plus spécifiquement, parmi les fondements théoriques convoqués dans l'étude du numérique en éducation figurent aussi le concept de « genèse instrumentale » (processus d'instrumentalisation/instrumentation) de Rabardel (1995), pour lequel « les techniques, les artefacts, les instruments sont, comme le langage ou les coutumes, constitutifs [du] milieu social », ainsi que le modèle du système général de l'activité d'Engeström (Engeström, 1987) et le « carré PADI[13] » de Wallet (2010) utilisé par Voulgre (2011, 2012).

Mais cette perspective de changements systémiques est intervenue dès la fin des années soixante, comme le constate Baron (2011) avec un « rapport de l'OCDE publié en 1971 [qui] fait le point sur la question de la technologie de l'éducation, conçue comme la mise en œuvre de nouveaux systèmes d'apprentissage (OCDE-CERI 1971) ». Baron et Depover (2019) justifient ainsi le recours à la notion d'écosystème et au cadre systémique « pour décrire la complexité des environnements d'apprentissage mobilisant des artefacts numériques (…) afin de rendre compte de la multiplicité et de l'interdépendance des variables qui conditionnent l'impact du dispositif pédagogique ».

Enfin, Collin et Brotcorne (2019) rapprochent cette ambition de l'approche critique « qui ne peut réduire les effets du numérique à une relation causale qu'il serait possible d'isoler et de mesurer » et Denouël (2019) voit dans cette approche sociotechnique une méthode pour ne pas isoler la variable technologique du tissu social dans ses dimensions matérielles, imaginaires et symboliques.

## 2.7 Une approche systémique du numérique centrée sur les savoirs

Cet état des connaissances montre donc que l'étude du numérique en éducation, complétée par les approches sociotechniques, systémiques et les apports des HN, recouvre des réalités multiples, sachant que notre grille de lecture et notre corpus (abordé spécifiquement ci-dessous) reflètent aussi notre itinéraire intellectuel et professionnel.

Mais comment ce changement plus global pour le monde des savoirs et de la *paideia*[14] peut-il être qualifié, notamment dans le cadre de l'histoire des sciences, des techniques et des savoirs?

En constatant que « notre rapport au savoir lui-même est modifié sous l'influence des technologies numériques », Baron et Depover (2019) identifient déjà plusieurs dimensions de cette modification : évolution des pratiques didactiques, accès et diffusion des savoirs, échanges et interactions.

Il peut être utile à cet effet de relier ces questionnements au concept de « régime de savoirs » établi par Pestre (2003), le nouveau régime de savoirs qu'il voit débuter à partir des années 1980 se caractérisant par une domination du « jeu scientifique » par des « (techno) sciences (…) orientées vers la production technologique » et par une recomposition des pratiques scientifiques « par le déploiement des outils informatiques et des banques de données » (Bonneuil *et al.*, 2015).

[13] Prenant en compte : « la Pédagogie, les Acteurs, le Dispositif, les Institutions et leurs effets interrelationnels » (Voulgre, 2011).

[14] Nous utilisons ce terme hérité du grec ancien, qui a l'avantage de combiner culture et corpus de connaissances contribuant à l'éducation et à la formation à la citoyenneté.

Nous proposons ici d'adapter et d'articuler ce concept avec les changements épistémologiques qui accompagnent la transformation numérique de la recherche, notamment ceux identifiés par Bachimont (2004) qui parle de « raison computationnelle » à l'instar de la raison graphique « qui qualifie la rationalité propre à l'écriture[15] » (Goody, 1979).

En nous recentrant sur la place des acteurs de l'éducation, de la formation et de la recherche, et les processus en cours dans ce nouveau régime de savoirs, quels peuvent être les choix conceptuels et méthodologiques pour construire une modélisation systémique du numérique en éducation?

# 3. Méthodologie

## 3.1 Recherche appliquée et réflexivité

Les choix conceptuels et méthodologiques présentés sont indissociables de notre démarche réflexive pour conduire nos actions de recherche appliquée. Cet effort de réflexivité se matérialise notamment par une organisation méthodologique et épistémologique spécifique de nos tâches professionnelles, incluant veille infodocumentaire, lectures, participation à des listes de diffusion et à des espaces d'échanges professionnels (notamment via les réseaux sociaux), gestion et alimentation de bases de données (présentées ci-dessous), participation à des colloques et séminaires, publications (Allouche, 2023b), formations, etc. En lien avec notre cadre théorique et ces choix, cette organisation s'est traduite principalement par un engagement dans le champ des HN pour comprendre de l'intérieur les implications sociales et épistémologiques de ces transformations, observées parallèlement de façon empirique dans nos pratiques professionnelles, dans leurs interconnexions avec l'éducation et la formation.

## 3.2 Choix conceptuels pour une approche systémique du numérique en éducation

Nos principaux choix conceptuels partent donc de ces fondements empiriques et théoriques, dont nous proposons ici une reprise selon deux axes : 1) apports des HN, 2) approche systémique, débouchant sur une proposition de synthèse.

1) Les apports des HN, dans le champ de la production des savoirs, résident dans une conception du numérique à la fois sociale et épistémique : favoriser la diversité des éclairages disciplinaires, inter- ou transdisciplinaires sur les pratiques numériques, donc prendre en compte la diversité des acteurs et des points de vue mobilisés, en couplant les apports de l'informatique et des sciences du numérique en tant que telles aux apports des recherches sur les modalités d'organisation, de production et de diffusion des savoirs issus des SHS. Pour comprendre les implications sur l'organisation et l'appropriation des savoirs, le numérique, étudié dans sa pervasivité, est aussi considéré comme culture et comme et milieu, dans ses implications instrumentales et épistémologiques. Ces apports ont eux-mêmes des implications méthodologiques concrètes comme la mise en œuvre systématique de l'accès ouvert pour les publications scientifiques et pédagogiques, produites avec l'appui et le financement des politiques publiques, ou le travail collaboratif dans la conduite des projets.

[15] Aux structures conceptuelles propres à la raison graphique (liste, tableau, formule), il met ainsi en regard celles de la nouvelle « raison computationnelle » (programme, réseau, couche). Bachimont (2017) évoque ainsi un nouveau « régime de connaissance » issu de la « rupture épistémologique » constituée par la « science des données », « le numérique [étant] à la fois une technologie intellectuelle et une ingénierie pour les systèmes physiques ».

2) Une approche systémique générale et sociocritique du numérique en éducation : considérer le numérique dans l'éducation comme un système dont l'entendement se conçoit d'abord en considérant la complexité des processus, des interactions et des rétroactions, intégrant les dimensions spécifiquement formelles et institutionnelles (programmes scolaires, évolution de la forme scolaire, politiques nationales et locales d'équipement, formations, etc.) et informelles (liens avec le numérique comme environnement social, pratiques non scolaires dans le cadre familial, loisirs, etc.). L'approche systémique rejoint en cela les travaux en anthropologie dans le sens d'une prise en compte globale, sociale et située des pratiques numériques.

En réponse à la tension toujours présente entre une essentialisation des technologies et une prise en compte des acteurs et des contextes de mise en œuvre, on peut s'appuyer sur l'une des composantes du pragmatisme (Hennion, 2015), en ne considérant pas les pratiques numériques en tant que faits sociaux, comme des choses figées, réifiées, mais comme des processus

À titre de synthèse, nous proposons donc de considérer le numérique dans ses multiples acceptions et dimensions, celles-ci reflétant des mises en tension qui sont autant de problèmes en suspens pour l'éducation et la recherche : à la fois comme technologie de l'écrit, culture, milieu, technologie intellectuelle et environnement sociotechnique qui réinvente sous nos yeux et de façon très rapide l'espace public, qui change notre représentation du monde et s'accompagne d'un nouveau rapport au savoir. En cela le numérique s'accompagne de mutations sociotechniques qui affectent l'espace de travail et d'expression de l'ensemble des « praticiens du savoir[16] » (du chercheur et du professeur à l'élève, en passant par le bibliothécaire, le conservateur de musée ou le médiateur culturel). La pervasivité du numérique débouche sur un nouveau régime de savoirs, incluant l'éducation et la recherche, qui s'accompagne de nouveaux questionnements éducatifs, didactiques et épistémologiques sur les modalités de production et de diffusion des savoirs, le rapport à la vérité, l'éducation à l'esprit critique.

## 3.3 Choix opérationnel de modélisation

Par modélisation, nous entendons ici une « action intentionnelle de construire, par composition de concepts et de symboles, des modèles susceptibles de rendre plus intelligible un objet ou un phénomène perçu complexe (…). En faisant fonctionner le modèle problème, on tente de produire des modèles-solutions » (Lugan, 2009). Pour Le Moigne (1994), la modélisation relève ainsi à la fois de l'acte de décision et d'instrumentation.

Par ce geste à la fois conceptuel et opérationnel, le cadre systémique se démarque en cela de cadres analytiques, comme celui exposé à l'intention des décideurs par OECD (2023), qui expose une approche holistique pour un écosystème politique d'éducation numérique, permettant d'identifier des leviers :

- à partir de six dimensions : 1) approches pédagogiques, curricula et évaluation, 2) gouvernance, orientation et réglementation, 3) financement et achat, 4) infrastructure et innovation, 5) renforcement des capacités, 6) politiques de ressources humaines;

- et sept niveaux d'analyse : 1) apprenants, 2) parents, 3) enseignants, 4) établissements, 5) autorités administratives régionales ou locales, 6) autorités administratives centrales, 7) autres acteurs de l'éducation numérique (notamment EdTech).

[16] Expression adaptée de Le Deuff (2017).

À ce titre, pour passer d'un cadre analytique à un cadre systémique, nous proposons comme stades intermédiaires les composantes de la transformation numérique de l'éducation (voir figure 2) et le modèle du système d'information dont nous proposons un schéma type centré sur les individus et les collectifs (voir figure 3).

**Figure 3**

*Modèle simplifié d'un système d'information*

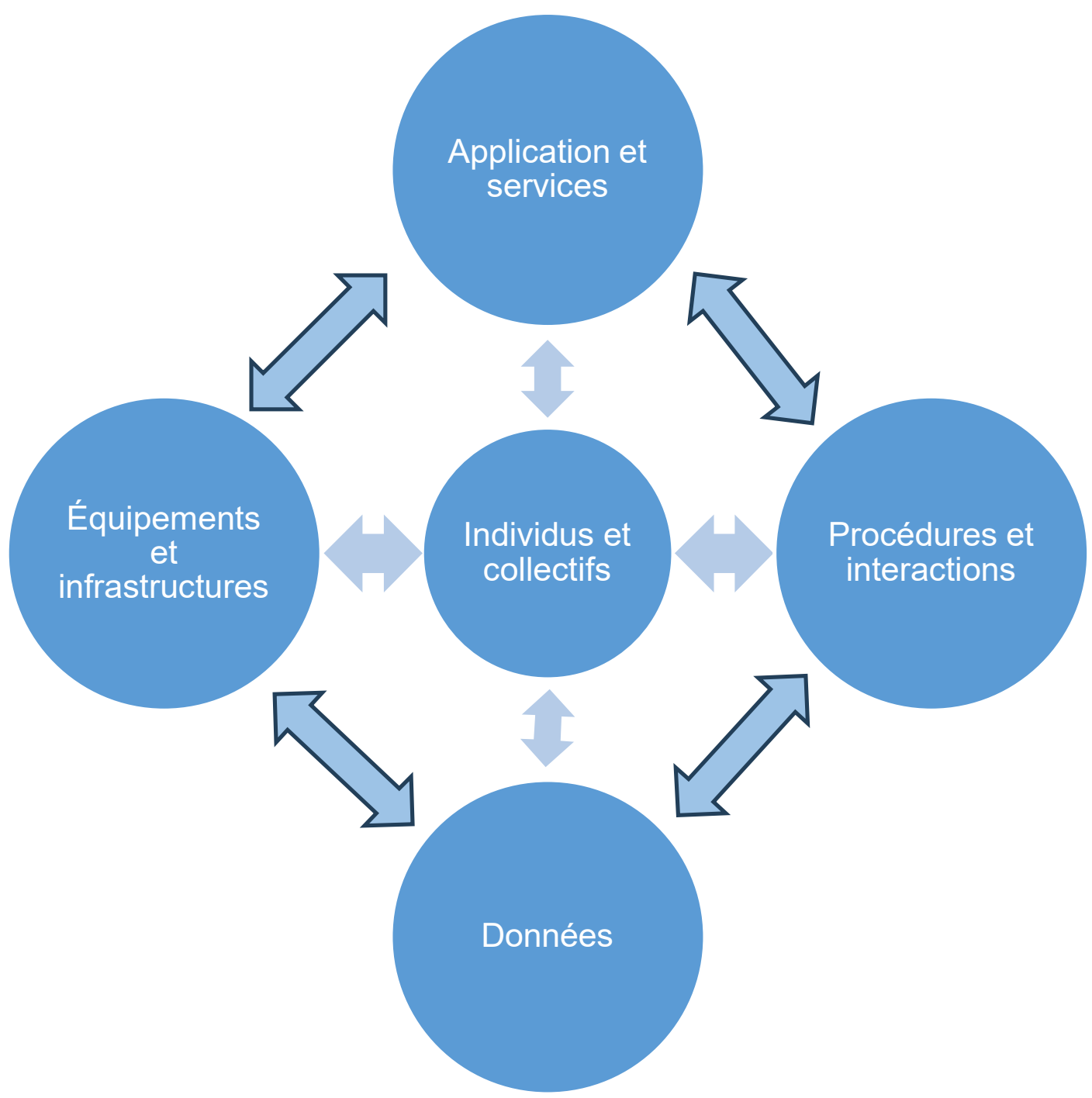

*Note*. © Elie Allouche.

Cette visualisation pourrait donc suggérer que le numérique en éducation peut être aussi représenté comme un système d'information, rejoignant en cela le besoin et la nécessité souvent exprimés dans la recherche comme dans les politiques publiques d'une éducation à la pensée informatique pour comprendre les procédures en jeu dans la transformation numérique et en tirer bénéfice pour l'éducation à la pensée critique, à la résolution de problème, à la collaboration ou à la créativité (Romero *et al.*, 2017), tout en ouvrant la voie à des enjeux et des pratiques pédagogiques non exclusivement informatiques ou technologiques.

Cependant elle laisserait de côté plusieurs des aspects sociaux, cognitifs, scientifiques, pédagogiques et formatifs qui sont pourtant au cœur des missions et des engagements professionnels des acteurs impliqués.

Les composantes de ce type de modèle sont donc appelées non pas à être ignorées, mais à prendre une place transversale et non structurelle dans notre modélisation.

Enfin, la démarche de modélisation est indissociable de notre rôle de modélisateur et revêt à ce titre un caractère performatif qui se traduit par des choix méthodologiques en matière de recherche appliquée.

## 3.4 Méthodologie des groupes thématiques numériques

Notre terrain d'application est le dispositif des GTnum (MENJ, 2023a) financé et coordonné par la DNE (Allouche, 2024a). Ce dispositif associe chercheurs (universités, laboratoires, groupements scientifiques) et praticiens (via les délégations régionales et académiques au numérique [DRANE/DANE] et des opérateurs de l'État comme le Réseau Canopé), favorisant ainsi la création de « collectifs hybrides de recherche » (Callon, 2012), l'accompagnement des pratiques numériques et leur appropriation professionnelle via la coproduction de savoirs en accès ouvert et l'objectivation des contextes (Plantard, 2023).

L'approche méthodologique que nous adoptons dans nos travaux de recherche appliquée met particulièrement l'accent sur deux dimensions essentielles : la recherche participative comme modalité de recherche appliquée consistant à conduire une recherche « par des consultations ou des implications à divers degrés » des acteurs de terrain (Renaud, 2020) et la science ouverte (Comité pour la science ouverte, 2021) consistant à favoriser systématiquement l'accès ouvert des données et publications issues de la recherche financée sur fonds publics. Cette orientation méthodologique se reflète concrètement dans l'espace de publication du carnet Hypothèses « Éducation, numérique et recherche » (DNE-TN2, 2019), dont nous sommes rédacteur en chef[17], soulignant ainsi l'importance de la diffusion des savoirs pour favoriser une transformation numérique de l'éducation ouverte et inclusive.

Par ailleurs, cette orientation accompagne la pratique, aide à formaliser et à mettre à distance, en intégrant une vision systémique dans la conception et la gestion des projets eux-mêmes, avec leurs implications professionnelles les plus concrètes dans la conception, l'organisation, la coordination et le suivi des groupes de travail :

- conception partenariale des axes thématiques avec les services, rectorats (services déconcentrés de l'État), opérateurs et partenaires du ministère;

- rédaction de l'appel à manifestation d'intérêt publié annuellement et communiqué sur les listes

---

[17] https://edunumrech.hypotheses.org/

de diffusion et les réseaux sociaux;

- réception et sélection des dossiers de candidatures constitués par les laboratoires porteurs, associés à des territoires académiques;

- conventionnement de trois ans entre le ministère (financeur du dispositif) et les universités de tutelle des laboratoires porteurs;

- réunions de cadrage méthodologique et de lancement;

- suivi des travaux et points d'étape;

- publication et valorisation en accès ouvert des travaux (états d'avancement, publications, plan de gestion de données et données de recherche), regroupés en portfolios (Allouche, 2024b);

- interventions lors de formations académiques et nationales, journées d'étude, séminaires, webinaires, etc.

Le dispositif des GTnum, aussi bien que celui des incubateurs académiques (MENJ, 2023c), relève donc d'une recherche appliquée et ouverte pour optimiser la diffusion des travaux et des productions, comme en témoigne l'organisation de notre corpus.

## 3.5 Corpus

Notre corpus est constitué des jeux de données présentées ci-dessous (BDD de 1 à 5), dont la collecte est issue d'un travail continu de gestion de projet, de coordination, de production et de traitement informationnel et documentaire. Le défi méthodologique et épistémologique est de transformer en données nos actions en combinant les dimensions quantitatives et qualitatives. Les données ainsi regroupées constituent un matériau réflexif et empirique, de la conception à la mise en œuvre.

Ce travail sur les données suppose de remettre en question les outils, environnements et supports, ainsi que les choix de catégorisations.

Les sources et bases de données associées à notre article peuvent être regroupées en plusieurs lots :

- BDD 1-Portfolio : données et documentation issues de l'itinéraire professionnel (Allouche, 2023b)

Regroupement, sélection et valorisation des principaux jalons et publications de notre parcours via le réseau social professionnel LinkedIn[18] et notre profil ORCID (*Open Researcher and Contributor ID*) avec la mise en réseau que ceux-ci impliquent autour de nos centres d'intérêt (numérique, humanités numériques, éducation, recherche, formation).

- BDD 2 : données et documentation issues des actions au sein de la DNE – les incubateurs académiques

[18] https://www.linkedin.com/in/elieallouche/

(MENJ, 2023c)

Productions issues des projets en académies associant chercheurs et praticiens de l'éducation au sein des structures d'incubation conduites par les délégations académiques ou régionales au numérique.

- BDD 3 : données et documentation issues des actions au sein de la DNE – les GTnum

    - BDD 3-Lot 1 (DNE-TN2, 2019) *passim*

    - BDD 3-Lot 2 (Allouche, 2024a)

Données quantitatives, thématiques traitées, partenaires mobilisés et accès aux productions.

- BDD 4 : laboratoires et acteurs scientifiques recensés à partir de la base (Ministère de l'Enseignement supérieur et de la Recherche, 2023)

Informations de référence permettant d'alimenter en continu la base de contacts des entités scientifiques travaillant sur le numérique en éducation ou dans des domaines voisins.

- BDD 5 : références bibliographiques issues des actions professionnelles et du travail de veille (Allouche, 2023a)

Cette base bibliographique et sitographique (16 547 références au 30/09/23) est construite selon une conception transversale du numérique. Centrée sur les HN, les modalités de production et de diffusion de savoirs en relation avec les questions d'éducation et de formation, elle se veut problématique dans son rubriquage, son alimentation et son évolution. Sans prétendre à une quelconque objectivité ni exhaustivité, elle reflète d'abord nos questionnements, nos difficultés et tentatives de résolution de celles-ci. Ainsi, plusieurs des références sont catégorisées dans plusieurs rubriques ou pourraient l'être en raison justement de cette transversalité. En prenant un exemple récent, l'accélération technologique des systèmes d'intelligence artificielle (IA) générative, dont les références récoltées relèvent, en première approche, de la rubrique « IA », a rapidement généré une sous-rubrique, tout en relevant potentiellement d'autres rubriques (pratiques pédagogiques, disciplinaires ou interdisciplinaires, littératie numérique, innovation ou prospective). Par ailleurs ce travail de veille nécessite d'assurer une couverture thématique très large, notamment sur les aspects sociaux et culturels, en raison des implications globales du numérique. À titre d'indicateur macro, nous présentons les cinq dossiers principaux et leur proportion au moment où ce texte est rédigé : 1) généralités (3,4 %), 2) recherche et pratique (50,8 %), 3) éducation (31,3 %), 4) culture, savoirs et société (10,3 %), 5) innovation (4,1 %). Les deux dossiers principaux (2 et 3) ont bien sûr de nombreuses passerelles thématiques, comme les questions méthodologiques (2.1), les littératies (2.5), la diffusion des savoirs (2.2.2), les communs (2.2.5), etc.

# 4. Résultats

Les résultats présentés visent à répondre à notre problématique et à mettre à l'épreuve nos hypothèses dans leurs dimensions théorique et applicative : complexité et transversalité du numérique, production de connaissances issue de la recherche appliquée via une modélisation systémique en appui aux politiques publiques.

## 4.1 Résultat 1 : Modélisation

Notre cadre théorique a identifié le constat d'une approche plurielle, hétérogène et complexe du numérique en éducation et d'une difficulté d'en dégager une synthèse pour les politiques publiques.

La modélisation systémique, réalisée à partir de notre cadre empirico-conceptuel, de nos choix méthodologiques et de l'exploitation de notre corpus, aspire donc à apporter une contribution à trois niveaux (voir figure 4) :

- replacer le numérique en éducation dans un environnement numérique plus large (comme donnée nodale du schéma), afin de prendre en compte la porosité entre les dimensions socioculturelles du numérique – éclairées par le champ des HN et les recherches en SHS – et le monde de l'éducation, évitant donc de poser le numérique comme « auto-agissant », hors contextes, intentions et décisions des acteurs;

- le numérique étant ainsi posé comme environnement global d'action-interaction-rétroaction, mettre en valeur les interactions et dynamiques collectives, qui ne relèvent pas exclusivement des politiques publiques, mais que celles-ci peuvent accompagner ou faciliter;

- prendre en compte les dynamiques communes entre recherche et éducation autour de l'hypothèse d'un nouveau « régime de savoirs », adapté de Pestre (2003), qui se déploierait à l'ère du numérique, afin d'intégrer les questions proprement scientifiques et didactiques (au-delà de l'évolutivité des dimensions spécifiquement technologiques). Corrélativement, comme l'attestent les nombreux témoignages et résultats d'enquêtes au moment de la pandémie de COVID (Administration & Éducation, 2021; DNE-TN2, 2021), il s'agit aussi de mettre en lumière les capacités des acteurs à inventer des solutions et à prendre des initiatives en réponse aux défis contemporains.

En se centrant sur les dynamiques de savoir, au cœur des finalités de l'éducation comme de la recherche, il s'agit aussi de représenter la complexité des phénomènes en jeu et leur imprévisibilité, correspondant à un champ de possibles, en rupture avec une approche technodéterministe.

Afin d'élargir notre approche, nous notons qu'une boucle itérative sociotechnique peut être identifiée en ces termes : le travail de modélisation doit intégrer le fait que les technologies, dans leur conception, incorporent des normes sociales, culturelles et politiques et qu'en retour leurs usages façonnent aussi des réalités sociales, culturelles, etc.

**Figure 4**

*Modélisation graphique de l'approche systémique du numérique en éducation*

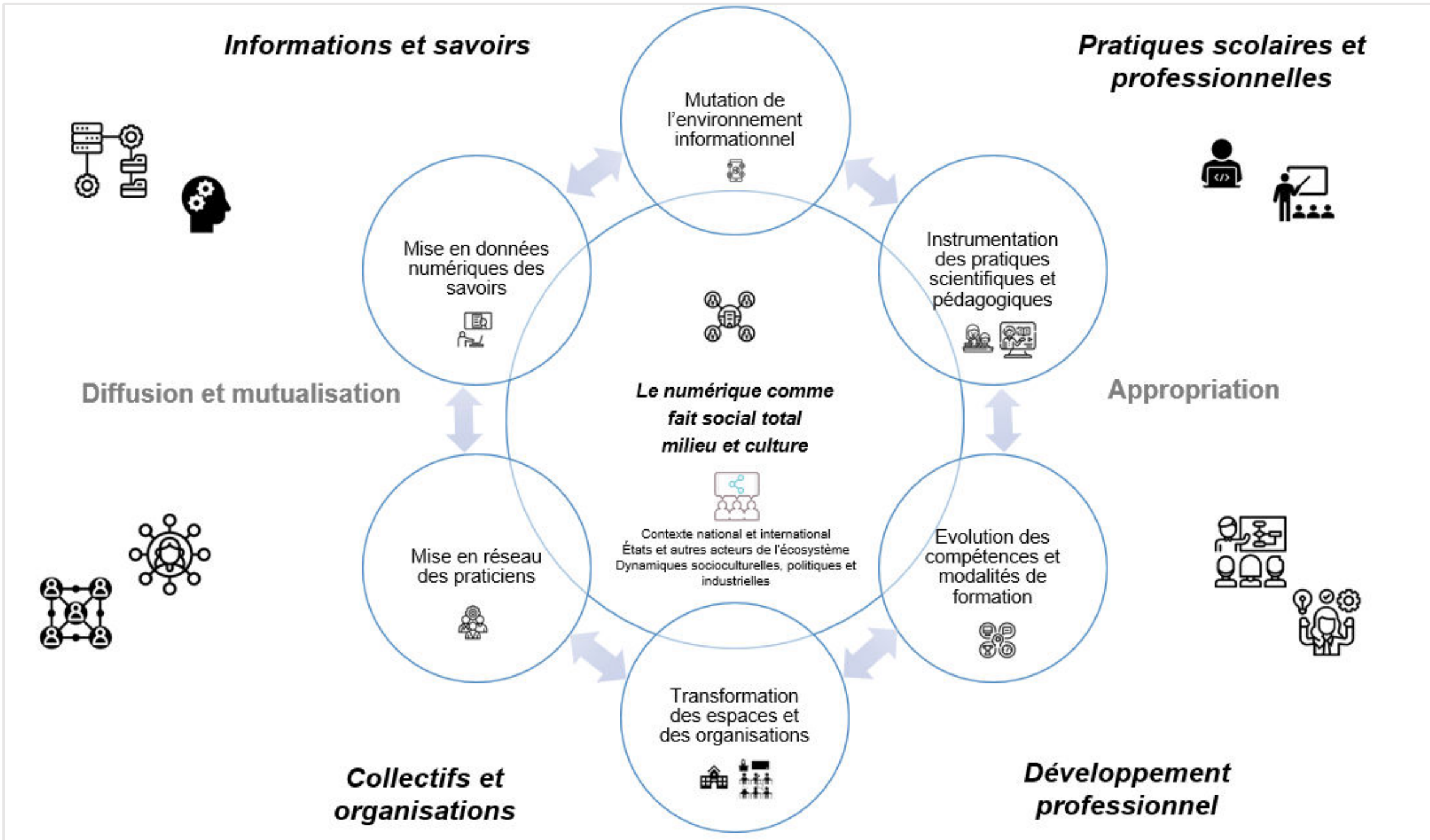

*Note*. © Elie Allouche.

Ainsi cette proposition de modélisation et de visualisation identifie :

- une donnée nodale : le numérique comme fait social total, milieu et culture;

Le numérique ainsi entendu constitue un champ de pratiques et un ensemble d'« épreuves », avec leur part d'indétermination, au sens où l'entend la sociologie pragmatique : « suivre au plus près la façon dont les acteurs s'engagent corporellement dans les dispositifs matériels qu'ils envisagent, ou qu'ils sont sommés de maîtriser » (Barthe *et al.*, 2013). Cette donnée nodale permet en outre de ne pas isoler les politiques publiques de leur contexte, national et international, des champs d'action et de la présence d'autres acteurs.

- quatre domaines d'actions-interactions-rétroactions : information et savoir, pratiques scolaires et professionnelles, développement professionnel, collectifs et organisations;

- deux dynamiques collectives centrales : diffusion et mutualisation et appropriation professionnelle;

- et six unités d’action, comme objets d’étude et de pratiques (dossiers 2 et 3 de notre BDD 5 et leurs passerelles thématiques), eux-mêmes interconnectés :

    1. Mutation de l’environnement informationnel;

    2. Instrumentation des pratiques;

    3. Évolution des compétences et modalités de formation;

    4. Transformation des espaces et des organisations;

    5. Mise en réseau des praticiens;

    6. Mise en données numériques des savoirs.

Cette visualisation ne doit pas masquer le fait qu’une modélisation est un processus dynamique et évolutif. Elle se comprend aussi par les flux qui concernent directement les politiques publiques : politiques nationales, académiques, initiatives locales, pratiques individuelles, collectives, communication et circulation d’informations, productions de contenus et par les articulations entre unités d’action et interactions. En s’inscrivant dans le cadre du numérique comme milieu et culture, il s’agit aussi de rappeler que cette culture numérique (Cardon, 2019) n’existe que par des individus et des collectifs, dotés d'agentivité à des degrés variables, qui effectuent des choix (par nature contextuels et discutables), agissent et interagissent.

Enfin, cette approche systémique est ouverte et non exhaustive. Ouverte, en s’intégrant elle-même dans plusieurs systèmes ou écosystèmes sous-jacents que sont principalement les établissements éducatifs, d’enseignement supérieur et de recherche, avec leurs propres systèmes d’information et rouages administratifs. Non exhaustive, en répondant d’abord à un besoin d’intelligibilité pour comprendre et agir dans un contexte en perpétuel changement, voire en changement accéléré. Elle est ouverte aussi en ce qu’elle ne préjuge pas des méthodes d’enseignement et de recherche mises en œuvre.

## 4.2 Résultat 2 : Test sur le corpus des groupes thématiques numériques

Le deuxième résultat propose l’articulation entre modélisation systémique et travaux des GTnum, publiés en accès ouvert sur un carnet de recherche (DNE-TN2, 2019; DNE-TN2, 2020; DNE-TN2, 2024) et répertoriés dans les jeux de données associés qui rendent compte des thématiques et des aspects quantitatifs (Allouche, 2024a). Il s’agit ainsi de reprendre les apports empirico-conceptuels, réinvestis dans les axes de modélisation, et les données produites par nos actions en recherche appliquée depuis leur mise en place en 2017.

Le premier niveau d’articulation se retrouve entre l’organisation des GTnum et la structure principale de la modélisation systémique :

- collectifs et organisations : mise en réseau de chercheurs et de praticiens via un espace de publication commun et des temps de rencontre (par exemple sur les aspects méthodologiques de la recherche participative ou des plans de gestion des données de recherche);

- informations et savoirs : publication en accès ouvert aux travaux et publications intermédiaires ou finales;

- pratiques professionnelles : accompagnement par la recherche de l'évolution des pratiques numériques en éducation;

- développement professionnel : dispositifs d'accompagnement et de formation en académie par la recherche;

- l'ensemble de ces actions est ainsi interconnecté avec la donnée nodale de la modélisation, en ce sens que les pratiques numériques qui se développent dans le cadre de ces groupes de travail se nourrissent et alimentent en même temps des pratiques quotidiennes, professionnelles ou pas.

Le deuxième niveau d'articulation (voir tableau 2) a pour objectif d'illustrer le caractère opératoire de notre modélisation dans la mesure où les apports de connaissances issus des GTnum sont ainsi affectés, non pas à des entrées systématiques (du type : matériel, infrastructures, contenus, etc.) qui atteignent leurs limites lorsqu'il s'agit d'identifier des leviers d'action caractéristiques d'un contexte professionnel, mais à des processus sociaux et professionnels replaçant l'éducation dans l'ensemble plus vaste des activités du savoir et dans lesquels le rôle et les initiatives des acteurs sont plus clairement identifiés. Via la donnée nodale d'un « numérique environnant », plutôt que réduit à ses dimensions technologiques (au risque de n'impliquer que les férus de technologie dans l'éducation), cette correspondance – et mise à l'épreuve interne au dispositif de recherche appliquée – permet aussi de mettre en avant l'interconnexion entre les processus à accompagner (par exemple, entre informations et savoirs et pratiques : instrumentation-environnement informationnel-mise en données des savoirs-évolution des compétences, lorsqu'il s'agit d'aborder la littératie des données dans les thématiques 2021-2024).

**Tableau 2**

*Correspondance entre axes de modélisation systémique et thématiques de recherche*

| Axes de modélisation systémique | Thématiques de recherche (exemples) |
|---|---|
| Collectifs et organisations – Transformation des espaces et des organisations | Nouveaux espaces d'apprentissage GTnum 2017-2020 |
| Pratiques scolaires professionnelles – Appropriation – Instrumentation des pratiques | Les usages numériques des jeunes GTnum 2017-2020 |
| Pratiques scolaires professionnelles – Appropriation – Instrumentation des pratiques | L'appropriation du numérique par les enseignants GTnum 2017-2020 |
| Collectifs et organisations – Mise en réseau des praticiens | Vers un renouvellement de l'écosystème numérique éducatif : acteurs, collectifs et organisations GTnum 2020-2022 |
| Focus sur la porosité, au moment de la crise sanitaire, entre donnée nodale et instrumentation des pratiques | Pour une prise en compte des disparités sociales et territoriales GTnum 2020-2022 |
| Focus sur la porosité, via les technologies d'IA, entre donnée nodale et instrumentation des pratiques | Intelligence artificielle et éducation GTnum 2020-2022 |
| Focus sur la porosité, au moment de la crise sanitaire, entre donnée nodale et instrumentation des pratiques des enseignants | Évolution des pratiques pédagogiques, des postures et des gestes professionnels des enseignants GTnum 2020-2022 |
| Informations et savoirs – Mise en données numériques des savoirs – Évolution des compétences | Littératie des données GTnum 2021-2024 |

*Note*. © Elie Allouche.

Sans pouvoir présenter dans le détail les résultats de chaque GTnum, on peut cependant noter que certains d'entre eux apportent un éclairage spécifique sur les dynamiques internes de notre modélisation : ainsi sur les usages et pratiques des jeunes et les conditions d'appropriation par les enseignants (DNE-TN2 et CREAD-M@rsouin, 2020a; DNE-TN2 et CREAD-M@rsouin, 2020b), la place croissante de l'IA (DNE-TN2 et Chaire RELIA, 2023; DNE-TN2 et LINE, 2023), avec une présentation de plusieurs initiatives françaises ou internationales en matière de formation ou de pratiques pédagogiques, ou sur les écosystèmes d'innovations dans les territoires (DNE-TN2 et TECHNÉ, 2023; Cerisier, 2024) débouchant sur des préconisations sur les conditions favorisant les « processus d'invention et d'innovation » (stabilité des politiques publiques, meilleure circulation de l'information, montée en compétences en ingénierie de projet).

# 5. Discussion

## 5.1 Rappel des objectifs

Nos objectifs étaient de comprendre la transformation numérique de l'éducation, en évaluant la capacité d'une recherche appliquée dans notre domaine professionnel de produire des connaissances dans un cadre d'étude empirico-conceptuel. Notre cadre théorique nous a conduit à produire une modélisation du numérique en éducation combinant les apports des HN, l'approche systémique et le concept de nouveau régime de savoirs. Nous avons présenté sa mise en application dans le cadre des politiques publiques, via le dispositif des GTnum, collectifs de recherche participative organisés avec le soutien de la DNE.

## 5.2 Résumé des principaux résultats

Les deux résultats proposés ont consisté d'une part en la visualisation de cette modélisation systémique du numérique en éducation, élargie aux domaines des pratiques du savoir (éducation, formation, recherche) dans le contexte d'un numérique considéré dans sa pervasivité socioculturelle; d'autre part dans le test de cette modélisation dans l'organisation et les productions des GTnum.

Ces résultats ont donc tenté de répondre à notre problématique par des apports théoriques et empiriques : théoriques en mobilisant l'état de la recherche sur le numérique (étude des technologies éducatives, apports des HN et approche systémique), en proposant de résoudre la tension existante entre essentialisation des technologies et prise en compte de la diversité des contextes et des modalités d'appropriation par les acteurs par une modélisation systémique plaçant comme donnée nodale un « numérique environnant »; empiriques en nous basant sur les faits d'expérience de notre itinéraire professionnel (enseignement, formation, direction d'établissement public, ingénierie de projet en recherche appliquée) pour constater que le numérique ne pouvait être réduit à ses dimensions technologiques et qu'il mettait en relation des processus complexes et une pluralité d'acteurs.

## 5.3 Interprétation

Nos travaux s'inscrivent dans la lignée des précédents cadres d'étude systémique et des approches sociotechniques et sociocritiques (présentés dans le cadre théorique) dans la mesure où il s'agit de reconnaître la complexité et la diversité des objets d'étude que recouvre le numérique en éducation dans ses dimensions socioculturelles.

La modélisation systémique proposée s'est donc voulue complémentaire des autres modélisations par ses apports interdisciplinaires combinant recherche sur le numérique en éducation et HN, en prenant en compte des questions très concrètes et pratiques, comme la pérennisation, voire la patrimonialisation des contenus et ressources scolaires, produites par les éditeurs ou par les enseignants eux-mêmes à titre individuel ou collectif, à l'image des initiatives prises depuis plusieurs années dans l'enseignement et la recherche avec la création et l'institutionnalisation de grandes infrastructures (Huma-Num, 2015) rejoignant par exemple la question de la professionnalisation des pratiques ou des communs en éducation (thématique d'un GTnum sur la période 2023-2026).

Nous proposons une interprétation de ces résultats à quatre niveaux.

### 5.3.1 COMPARAISON AVEC D'AUTRES APPROCHES SYSTÉMIQUES

Comparée au « carré PADI » de (Wallet, 2010), notre modélisation retient l'interconnexion entre les quatre pôles considérés [pédagogie-acteurs-dispositifs-institution], en proposant d'une part une prise en compte accrue du contexte socioculturel tel qu'il s'est transformé depuis les années 2000 avec un numérique « ambiant » et « pervasif » (Delmas-Rigoutsos, 2018; Boullier, 2016), concernant à ce titre l'ensemble des secteurs d'activité et des objets d'étude, d'autre part en cherchant à identifier des enjeux et processus (environnement informationnel, mise en données des savoirs, transformation des espaces).

Comparés au cadre sociocritique de Collin et Brotcorne (2019), nos choix retiennent le fait de « ne [pas] réduire les effets du numérique à une relation causale » et l'existence de « rapports sociaux multiples, multidirectionnels », en élargissant cette prise en compte aux domaines du savoir pour intégrer par exemple la nécessité de compétences spécifiques, comme celles liées aux données numériques, compétences déjà requises dans la recherche et les métiers de l'information, de la communication et de la documentation.

### 5.3.2 AXES DE MODÉLISATION ET ÉVOLUTION DES PRATIQUES

Nous proposons ici une mise en correspondance complémentaire à partir des enjeux de l'accompagnement de la transformation numérique de l'éducation tels qu'ils sont identifiés dans les feuilles de route pour la politique numérique de territoires académiques : à titre d'exemple dans Académie de Bordeaux (2021) et Région académique Guyane (2023). Les quelques axes stratégiques présentés dans la colonne de droite illustrent la priorité donnée aux dynamiques collectives autour des pratiques, de la formation ou des contenus éducatifs. Il est ainsi possible de prolonger ces axes vers d'autres enjeux liés à la porosité entre les unités d'action identifiées dans la modélisation : par exemple entre l'instrumentation des pratiques, l'évolution des compétences et la transformation des espaces et organisations (à l'échelle d'un territoire ou d'un établissement scolaire).

**Tableau 3**

*Correspondance entre axes de modélisation systémique et accompagnement de la transformation numérique*

| Axes de modélisation systémique | Axes de feuille de route académique (exemples) |
|---|---|
| Pratiques scolaires professionnelles – Appropriation – Instrumentation des pratiques | Exploiter le numérique comme valeur ajoutée des pratiques d'apprentissages et d'enseignements |
| Collectifs et organisations – Mise en réseau des praticiens | Consolider l'écosystème dans le sens d'une politique partagée<br>Créer un environnement propice au déploiement du numérique dans l'ensemble du système éducatif régional |
| Développement professionnel – Évolution des compétences et modalités de formation | Développer l'enseignement numérique et les compétences numériques<br>Consolider les compétences de tous les acteurs |
| Informations et savoirs – Diffusion et mutualisation – Mise en réseau de praticiens | Poursuivre l'outillage des utilisateurs et le déploiement des communs numériques |

*Note*. © Elie Allouche et Académie de Bordeaux (2021) et Région académique Guyane (2023).

### 5.3.3 PERSPECTIVES MÉTHODOLOGIQUES

Concernant notre cadre méthodologique de recherche appliquée et de recherche participative, même si la place de la recherche dans la formation initiale et continue des enseignants fait l'objet de plusieurs publications, espaces et notes de veille (comme ceux de l'Institut français de l'éducation et de Réseau Canopé[19]), il faut reconnaître que l'impulsion doit s'inscrire dans la durée, comme l'indiquent les recommandations du récent rapport (IH2EF, 2023) pour viser « un processus global de transfert entre recherche et pratique dans le scolaire ».

En revanche, via le dispositif présenté ici, selon nous il ne s'agit pas tant, ni exclusivement, d'un « transfert » vers les praticiens que d'un processus de coconstruction de connaissances à inventer ou à réinventer pour traiter des situations scientifiques et éducatives dans leur contexte et leur complexité. La question demeure aussi de la prise en compte des seules données dites « probantes » issues de la recherche pour accompagner les enseignants et guider les politiques publiques, ou d'une plus large variété de données issues de la pratique professionnelle et intégrant la diversité des points de vue et des contextes d'application (Allaire *et al.*, 2023; Baron et Fluckiger, 2021). Parmi les pistes de travail figurent aussi la production d'« objets frontières » facilitant le dialogue entre chercheurs et praticiens (Monod-Ansaldi *et al.*, 2019)[20].

Nous pouvons aussi évoquer les différences de temporalité existant entre temps de la recherche (a minima 3-5 ans), temps d'application, notamment en matière d'enseignement et de formation (années scolaires, cycles d'études, conception et application des programmes et référentiels), et temps des politiques

---

[19] https://veille-et-analyses.ens-lyon.fr/ et https://www.reseau-canope.fr/agence-des-usages.html

[20] « L'objet frontière s'entend comme un dispositif permettant d'amorcer un travail commun entre plusieurs mondes et assurant une flexibilité suffisante pour que chaque acteur puisse trouver un intérêt à son étude ou à son usage » (Monod-Ansaldi *et al.*, 2019).

publiques (calendrier politique et législatif, prise en compte des enjeux internationaux).

Le pari pour les politiques publiques est donc de s'appuyer à la fois sur des savoirs scientifiques et expérientiels, sur des données quantitatives et qualitatives (Ifé, 2022) tout en participant – et notre proposition de modélisation se veut illustrative à ce propos – à l'effort collectif d'intelligibilité et d'action.

#### 5.3.4 ÉLÉMENTS DE PROSPECTIVE

Comme indiqué en introduction, notre modélisation a aussi pour objectif de poser un cadre non pas statique, mais à vocation prospective, en prenant comme exemple un thème au cœur de l'actualité, celui du développement des systèmes d'IA générative et de leur impact à venir.

À partir des six unités d'action identifiées dans notre approche systémique, les tendances observées s'appuient notamment sur nos publications sur le thème de l'IA et de l'éducation, notamment Allouche (2023c) et DNE-TN2 (2023). Nous observons ainsi que ces tendances concernent surtout, mais non exclusivement, les trois premières unités d'action relatives à l'environnement informationnel, aux pratiques et aux compétences.

**Tableau 4**

*Modélisation systémique et prospective sur l'IA générative*

| **Modélisation systémique : unités d'action** | **Tendances émergentes liées à l'IA générative** |
|---|---|
| 1. Mutation de l'environnement informationnel | Développement de systèmes IA multimodaux intégrés à l'environnement informationnel, permettant de produire et/ou d'adapter des contenus à des publics et des besoins multiples.<br>Renouvellement des modalités d'éducation, de formation et d'évaluation, nécessité accrue d'éducation à l'esprit critique à tous les niveaux.<br>Défi pour l'éducation aux médias et à l'information. |
| 2. Instrumentation des pratiques | Généralisation des IA génératives dans l'expérience utilisateur des principales suites bureautiques, applications, plateformes et réseaux sociaux. |
| 3. Évolution des compétences et modalités de formation | Nouvelles compétences émergentes et montée en compétences dans les organisations intégrant l'IA dans le tronc commun et les domaines de spécialité.<br>Couplage entre ingénierie des instructions/requêtes (*prompting*) et ingénieries métiers. |
| 4. Transformation des espaces et des organisations | Fusion entre IA génératives, réalité augmentée et environnements immersifs. |
| 5. Mise en réseau des praticiens | Modalités d'organisation et de répartition du travail intégrant des systèmes IA de plus en plus performants, voire autonomes, dès l'amont des projets et les processus. |
| 6. Mise en données numériques des savoirs | Croisement des littératies (numérique/informationnelle, des données, en IA). |

*Note*. © Elie Allouche.

## Conclusion

Nous avons souhaité proposer dans cet article la construction d'une approche systémique du numérique en éducation centrée sur les savoirs pour intégrer et concilier la multiplicité des objets et des dimensions que recouvre le numérique en éducation, en articulation avec les apports d'une recherche appliquée via notre terrain d'application professionnelle et le dispositif de recherche participative des GTnum.

Dans un premier temps, nous avons posé les bases d'un cadre théorique intégrant un état général des connaissances sur le numérique en éducation, complété par les apports des HN et de l'approche systémique comme réponse à la tension constatée entre essentialisation des technologies, dans leurs effets positifs ou négatifs sur les apprentissages, et prise en compte de la complexité des contextes et de la place du choix des acteurs dans la réussite ou l'échec d'un dispositif sociotechnique. En réponse à ce constat, nous avons identifié un « numérique environnant » et transversal, se traduisant comme donnée nodale dans notre proposition de modélisation systémique, celle-ci pouvant être posée comme cadre d'intelligibilité pour les politiques publiques et testée à ce titre dans le cadre des GTnum ou de l'accompagnement de la transformation numérique dans les territoires académiques. La méthodologie utilisée relève d'une démarche réflexive articulée à des choix conceptuels (les HN pour élargir aux dimensions socioculturelles du numérique appliquées à un nouveau régime de savoirs et l'approche systémique pour la prise en compte de la complexité des processus en jeu) et à la modalité de recherche participative associant recherche et praticiens de l'éducation.

Étant nous-mêmes praticien engagé, produisant et appliquant cette modélisation comme cadre d'intelligibilité et d'action pour les politiques publiques, nous avons conscience que cette posture constitue elle-même un biais dans notre grille de lecture et nos choix méthodologiques. Néanmoins, au terme de cette étude, notre conclusion est que le numérique dans l'éducation peut se concevoir comme une entité complexe, plurielle et polymorphe : ce ne sont pas tant les technologies en elles-mêmes qui constituent nos objets d'étude ou cadres d'action qu'une pluralité de situations, d'acteurs et d'interactions, combinant l'institutionnel, l'informel, l'ancien et le nouveau.

Par ailleurs, l'une des limites de l'approche systémique et de son application est l'acceptation encore très diverse du numérique dans l'éducation ou la recherche. Une tension demeure en effet entre la compréhension des enjeux (variable selon les acteurs et les contextes), l'intégration inégale, ponctuelle ou discontinue des apports de la recherche (par exemple lors des temps de formation) et la nécessité d'actions opérationnelles dans les politiques publiques, l'accompagnement, la formation ou la mise en œuvre de services.

Enfin, parmi les implications envisagées et les recherches futures à mener peut figurer l'anticipation d'interactions croissantes entre systèmes d'IA générative et dispositifs d'ingénierie pédagogique ou de formation, l'une de nos hypothèses étant que s'ébauche actuellement une nouvelle configuration du numérique comme milieu dans lequel les savoirs et les situations d'apprentissage résulteraient d'hybridations entre humains et technologies d'IA, celles-ci étant elles-mêmes les productions et les extensions de l'intelligence humaine appliquée à la production, à la diffusion des savoirs et à la créativité.

# Liste de références

Académie de Bordeaux (2021). *Feuille de route pour le numérique éducatif dans la région académique Nouvelle-Aquitaine*. DRANE Nouvelle-Aquitaine. http://tinyurl.com/3zemzhub

Administration & Éducation (2021). École et crise sanitaire : déstabilisation et opportunités. *Administration & Éducation*, *169*. https://www.cairn.info/revue-administration-et-education-2021-1.htm

Aillerie, K. (2017). Le « numérique éducatif » à l'épreuve des pratiques scolaires : petits arrangements avec le marché. *Hermès, La Revue Cognition, communication, politique*, (78). https://doi.org/10.3917/herm.078.0023

Allaire, S., Granger, N., Tremblay, M., et Leroux, M. (2023, janvier 23). *Au-delà des données probantes : l'importance de la diversité des connaissances issues des recherches en éducation*. CTREQ - RIRE. http://tinyurl.com/2mkpzct9

Allouche, E. (2020, septembre). Les humanités numériques, pour un dialogue interdisciplinaire entre recherche et éducation. *Frantice.net*, (17), 59-69. http://frantice.net/docannexe/file/1685/frantice_17_.pdf

Allouche, E. (2022, 5 décembre). Numérique éducatif et recherche appliquée : les actions de la DNE. *Adjectif.net.* https://adjectif.net/spip.php?article575

Allouche, E. (2023a). Humanités numériques, éducation et formation. Bibliothèque Zotero [Bibliothèque Zotero]. https://www.zotero.org/groups/228138/humanits_numriques_ducation_et_formation/collections/QSEBUZXI

Allouche, E. (2023b). Liste de publications (ORCDI 0000-0001-8015-8198) [jeu de données]. HAL Science ouverte. https://cv.hal.science/elieallouche

Allouche, E. (2023c, octobre 2). IA génératives et ingénierie pédagogique : le prompting, pistes de travail et applications [Billet]. *Éducation, numérique et recherche*. https://edunumrech.hypotheses.org/9934

Allouche, E. (2024a). Liste, travaux et publications des Groupes thématiques numériques (2017-2023) [jeu de données]. Zenodo. https://doi.org/10.5281/zenodo.10521557

Allouche, E. (2024b, janvier 4). Groupes thématiques numériques 2020-2022 : portfolios [Billet]. *Éducation, numérique et recherche*. https://edunumrech.hypotheses.org/1076 8

Allouche, E., et Desfriches-Doria, O. (2021). Cartographie et enjeux des formations en Humanités Numériques. Dans P. Bonfils et J. Walter, *Questionner les humanités numériques* (p. 227-257). SFSIC et CPdirsic. https://www.sfsic.org/publications-sfsic/ouvrages-actes/questionner-les-humanites-numeriques/

Althaus, V., Grosjean, V., et Brangier, É. (2013). La centration sur le processus du changement : L'apport de l'intervention systémique à l'amélioration du bien-être au travail. *Activités*, *10*(1). https://doi.org/10.4000/activites.607

Azoulay, B., et De Courson, B. (2021). *Gallicagram : un outil de lexicométrie pour la recherche* [Prépublication]. SocArXiv. https://doi.org/10.31235/osf.io/84bf3

Bachimont, B. (2004). Arts et sciences du numérique : ingénierie des connaissances et critique de la raison computationnelle. *Mémoire de HDR*. http://tinyurl.com/4pbm57ad

Bachimont, B. (2017). Le numérique comme milieu : enjeux épistémologiques et phénoménologiques : principes pour une science des données. *Interfaces numériques*, *4*, 237ko. https://doi.org/10.25965/interfaces-numeriques.386

Baron, G.-L. (1994). *L'informatique et ses usagers dans l'éducation* [Habilitation à diriger des recherches, Université René Descartes - Paris V]. https://tel.archives-ouvertes.fr/edutice-00000370

Baron, G.-L. (2011). « Learning design ». *Recherche et formation*, *68*. https://doi.org/10.4000/rechercheformation.1565

Baron, G.-L. (2014). Élèves, apprentissages et « numérique » : Regard rétrospectif et perspectives. *Recherches en éducation*, *18*, 91-103. http://www.mutatice.net/glbaron/lib/exe/fetch.php/baron_article_oct_13_revu.pdf

Baron, G.-L. (2020). Brèves réflexions sur les humanités numériques. *Frantice.net*. http://mutatice.net/glbaron/lib/exe/fetch.php/hn_baron_frantice_v3.pdf

Baron, G.-L., Bounay, M., Dautrey, P., Guelfucci, J., Hebert, D., Muller, P., Schwob, M., et Tourtelier, P. (1981). *Dix ans d'informatique dans l'enseignement secondaire*. INRP. https://www.epi.asso.fr/blocnote/Dix_ans_INRP_1981.pdf

Baron, G.-L., et Boulc'h, L. (2011). *Les technologies de l'information et de la communication à l'école primaire. État de question en 2011*. EPI. http://www.epi.asso.fr/revue/articles/a1202b.htm

Baron, G.-L., Wallet, J., et Voulgre, E. (2018, 14 juin). Technologies de l'information et de la communication et Sciences de l'éducation. *Adjectif*. https://adjectif.net//spip.php?article468

Baron, G.-L., et Depover, C. (dir.). (2019). *Les effets du numérique sur l'éducation : regards sur une saga contemporaine*. Presses universitaires du Septentrion.

Baron, G.-L., et Fluckiger, C. (2021). Approches et paradigmes pour la recherche sur les usages éducatifs des technologies. Enjeux et perspectives. *Revue canadienne de l'apprentissage et de la technologie*, *47*. https://hal.archives-ouvertes.fr/hal-03349957

Barthe, Y., de Blic, D., Heurtin, J.-P., Lagneau, É., Lemieux, C., Linhardt, D., Moreau de Bellaing, C., Rémy, C., et Trom, D. (2013). Sociologie pragmatique : Mode d'emploi. *Politix*, *103*(3), 175-204. https://doi.org/10.3917/pox.103.0173

Basque, J., et Lundgren-Cayrol, K. (2002). Une typologie des typologies des applications des TIC en éducation. *Sciences et Technologies de l'Information et de la Communication pour l'Éducation et la Formation*, *9*(3), 263-289. https://doi.org/10.3406/stice.2002.1510

Bechetti-Bizot, C. (2017, mai). *Repenser la forme scolaire à l'heure du numérique : vers de nouvelles manières d'apprendre et d'enseigner (rapport de l'Inspection générale de l'Éducation nationale)*. Ministère de l'Éducation nationale et de la Jeunesse. http://tinyurl.com/3v7dz6c3

Bonneuil, C., Pestre, D., Breteau, C., et Le Roy, C. (2015). *Le siècle des technosciences, depuis 1914*. Éditions du Seuil.

Boullier, D. (2016). *Sociologie du numérique*. A. Colin.

Bourdeloie, H. (2014). Ce que le numérique fait aux sciences humaines et sociales. Épistémologie, méthodes et outils en questions. *tic&société*, *7(2)*. https://doi.org/10.4000/ticetsociete.1500

Bruillard, É. (1997). *Les machines à enseigner*. Hermès.

Bruillard, É., Baron, G.-L., Mendelson, P., Lombard, F., Pelgrims-Ducrey, G., et Coutret, G. (2002). Pratiquer les TICE. Chapitre 4. Compétences requises. Dans R. Guir (dir.), *Pratiquer les TICE* (p. 255-289). De Boeck Supérieur. https://www.cairn.info/pratiquer%20les%20tice--9782206082110-page-255.htm

Callon, M. (réalisateur) (2012, 3 octobre). *Les sciences sociales confrontées aux nouvelles pratiques de recherche et d'innovation* [vidéo]. YouTube https://www.youtube.com/watch?v=ZvJyAE6X3Z0

Cantrel, C. (2018, avril 19). OpenEdition, « un cas à part dans la littérature académique dédiée à l'Open Access » [Billet]. *L'Édition électronique ouverte*. https://leo.hypotheses.org/13752

Cardon, D. (2019). *Culture numérique*. Les presses SciencesPo.

Carlin, M., et Laborderie, A. (2021). Le BnF DataLab, un service aux chercheurs en humanités numériques. *Humanités numériques*, *4*. https://journals.openedition.org/revuehn/2684

Ceci, J.-F. (2018, février). *Pourquoi le numérique éducatif fait-il tant débat autour des bénéfices que l'on peut en attendre ? Explications via la métaphore de l'amplificateur pédagogique et définition de la pédagogie à l'ère du numérique.* ResearchGate. http://doi.org/10.13140/RG.2.2.21636.22400

Cerisier, J.-F. (2015). *La forme scolaire à l'épreuve du numérique*. https://hal.archives-ouvertes.fr/hal-01216702

Cerisier, J.-F. (2024, janvier 16). Une école en « transition numérique », vraiment? *The Conversation.* http://theconversation.com/une-ecole-en-transition-numerique-vraiment-220733

CNESCO (2020). *Numérique et apprentissages scolaires.* Conseil national d'évaluation du système scolaire. https://www.cnesco.fr/numerique-et-apprentissages-scolaires/

Collin, S. (2023, juillet 3). *La conception des technologies éducatives comme configuration des inégalités socio-numériques d'usage* (INSPE de Bretagne, Bretagne, France) [Video]. Pod Inspé Bretagne; INSPE de Bretagne. http://tinyurl.com/2smrkw8m

Collin, S., et Brotcorne, P. (2019). Contribution d'une approche sociocritique à l'étude des effets du numérique en éducation. Dans G. L. Baron et C. Depover (dir.), *Les effets du numérique sur l'éducation. Regards sur une saga contemporaine* (p. 229-243). Septentrion.

Comité pour la science ouverte (2021, juillet). Deuxième Plan national pour la science ouverte. *Ouvrir la Science*. https://www.ouvrirlascience.fr/deuxieme-plan-national-pour-la-science-ouverte

Dacos, M. (2010). Manifeste des Digital humanities (humanités numériques) [Billet]. *THATCamp Paris*. http://tcp.hypotheses.org/318

Dacos, M., et Mounier, P. (2014). *Humanités numériques : état des lieux et positionnement de la recherche française dans le contexte international* (Centre; Institut français). http://tinyurl.com/3h3eb8yd

Dariah Clarin (2014, 2021). *Digital Humanities Course Registry*. Digital Humanities Course Registry. https://dhcr.clarin-dariah.eu/courses/

Davallon, J. (2004). Objet concret, objet scientifique, objet de recherche. *Hermès, La Revue*, *38*(1), 30-37. https://www.cairn.info/revue-hermes-la-revue-2004-1-page-30.htm

De Lavergne, C. (2007*). La posture du praticien-chercheur : un analyseur de l'évolution de la recherche qualitative*. *Recherches qualitatives,* HS 3. http://www.recherche-qualitative.qc.ca/documents/files/revue/hors_serie/hors_serie_v3/Delavergne-FINAL2.pdf

Delmas-Rigoutsos, Y. (2018, 7 février). *Proposition de structuration historique des concepts de la pensée informatique fondamentale*. Didapro 7 – DidaSTIC. De 0 à 1 ou l'heure de l'informatique à l'école. HAL. https://hal.archives-ouvertes.fr/hal-01752797

Denouël, J. (2019). D'une approche sociocritique à une approche sociotechnique critique des usages numériques en éducation. *Formation et profession*, *27*(3), 36-48. http://doi.org/10.18162/fp.2019.483

Diemer, A. (2014). L'éducation systémique, une réponse aux défis posés par le développement durable. *Éducation relative à l'environnement. Regards - Recherches - Réflexions*, *11*. https://doi.org/10.4000/ere.805

DNE-TN2 et Chaire RELIA (2023, 6 septembre). Intelligence artificielle et éducation ouverte : portfolio du GTnum #IA_EO [Billet]. *Éducation, numérique et recherche*. https://edunumrech.hypotheses.org/9781

DNE-TN2 et CREAD-M@rsouin (2020a, 27 mars). Pratiques et usages numériques des jeunes : productions du GTnum 4 [Billet]. *Éducation, numérique et recherche*. https://edunumrech.hypotheses.org/1429

DNE-TN2, et CREAD-M@rsouin. (2020b, 6 avril). Les enseignants et le numérique : productions du GTnum 9 [Billet]. *Éducation, numérique et recherche*. https://edunumrech.hypotheses.org/1497

DNE-TN2 et LINE (2023, 10 juillet). Enseigner et apprendre à l'ère de l'intelligence artificielle : portfolio du GTnum #Scol_IA [Billet]. *Éducation, numérique et recherche*. https://edunumrech.hypotheses.org/9593

DNE-TN2 et TECHNÉ (2023, 20 décembre). Numérique scolaire : vers des écosystèmes favorables à l'innovation : portfolio du GTnum TECHNE #REVE [Billet]. *Éducation, numérique et recherche*. https://edunumrech.hypotheses.org/10652

DNE-TN2 (2019). Éducation, numérique et recherche : veille et diffusion des travaux de recherche sur le numérique dans l'éducation. *Éducation, numérique et recherche*. https://edunumrech.hypotheses.org/

DNE-TN2 (2020, 3 juillet). Productions des groupes thématiques numériques de la DNE (2017-2020) [Billet]. *Éducation, numérique et recherche*. https://edunumrech.hypotheses.org/1948

DNE-TN2 (2021, 26 janvier). École, numérique et confinement : situation à l'international et état la recherche en France (visuels bilingues) [Billet]. *Éducation, numérique et recherche*. https://edunumrech.hypotheses.org/2602

DNE-TN2 (2023, 9 mai). Intelligence artificielle et éducation : apports de la recherche et enjeux pour les politiques publiques [Billet]. *Éducation, numérique et recherche*. https://edunumrech.hypotheses.org/?p=8726

DNE-TN2 (2024, 4 janvier). Groupes thématiques numériques 2020-2022 : portfolios [Billet]. *Éducation, numérique et recherche*. https://edunumrech.hypotheses.org/10768

Doueihi, M. (2013). *Qu'est-ce que le numérique?* Presses universitaires de France. https://www.cairn.info/qu-est-ce-que-le-numerique--9782130627180.htm

Ellul, J. (1977). *Le système technicien*. Calmann-Lévy.

Engeström, Y. (1987). *Learning by expanding: An activity-theoret. approach to developmental research*. Orienta-Konsultit Oy. https://lchc.ucsd.edu/MCA/Paper/Engestrom/expanding/toc.htm

ENS PSL. (2021). *Observatoire des humanités numériques*. Observatoire des humanités numériques. https://odhn.ens.psl.eu/article/que-sont-les-humanites-numeriques

Fluckiger, C. (2016). Culture numérique, culture scolaire : homogénéités, continuités et ruptures. *Diversité, VEI*, *185*, 64-70. http://hal.univ-lille3.fr/hal-01588410

Fluckiger, C. (2017). Les technologies numériques à l'école, quel bilan? Dans F. Thibault et C.Garbay (dir.), *La recherche sur l'éducation*, vol. 2, Contribution des chercheurs, rapport remis à M Thierry *Mandon* (p. 103-104) [Rapport de recherche]. Ministère de l'Éducation nationale. http://hal.univ-lille3.fr/hal-01613680

Fluckiger, C. (2019). *Du dispositif à l'environnement : Le déterminisme technique à l'aune de l'évolution des usages estudiantins Cédric Fluckiger* (p. 368). Éditions Raisons et passions.

Fluckiger, C. (2020). *Les usages effectifs du numérique en classe et dans les établissements scolaires*. CNESCO. http://www.cnesco.fr/wp-content/uploads/2021/02/210218_Cnesco_Fluckiger_Numerique_Usages.pdf

Gille, B. (1978). *Histoire des techniques : technique et civilisations, technique et sciences*. Galimard.

Goody, J. (1979). La raison graphique : la domestication de la pensée sauvage. Les Éditions de Minuit.

Hardouin, M., Keerle, R., Thémines, J.-F., Boudesseul, G., Danic, I., David, O., Fontar, B., Guibert, C., Guillemot, L., Mentec, M. L., Plantard, P., et Plouchart-Even, L. (2018). Des fonctions d'un glossaire dans un programme de recherche pluridisciplinaire. *EspacesTemps.net Revue électronique des sciences humaines et sociales.* http://tinyurl.com/4d559e6h

Hcérès (2020). Référentiel d'évaluation des unités de recherche : campagne d'évaluation 2020-2021 vague b. Haut conseil de l'évaluation de la recherche et de l'enseignement supérieur. https://www.hceres.fr/sites/default/files/media/downloads/referentiel_ur_vague-b_rech-ur_22juin2020.pdf

Hennion, A. (2015). Enquêter sur nos attachements. Comment hériter de William James? *SociologieS*. https://doi.org/10.4000/sociologies.4953

Humanistica (2020). Humanités numériques (revue éditée par l'association Humanistica). *Humanités numériques*. http://journals.openedition.org/revuehn

Huma-Num (2015, mars 24). *Huma-Num : l'infrastructure des humanités numériques*. https://www.huma-num.fr/quest-ce-que-l-ir-huma-num/

Ifé (2022, septembre 14). *#44 La preuve en éducation : la WebRadio de l'Institut Français de l'Éducation*. Institut Français de l'Éducation (Ifé). http://ife.ens-lyon.fr/kadekol/ca-manque-pas-dr/44-la-preuve-en-education

IH2EF (2023). *L'École dans une société numérique*. IH2EF. https://www.ih2ef.gouv.fr/lecole-dans-une-societe-numerique

Inaudi, A. (2017a). École et numérique : quelques dates clés. *Hermès, La Revue*, *78*, 19-22. http://www.cairn.info/revue-hermes-la-revue-2017-2-p-19.htm

Inaudi, A. (2017b). École et numérique : une histoire pour préparer demain. *Hermès, La Revue*, *78*, 72-79. http://www.cairn.info/revue-hermes-la-revue-2017-2-p-72.htm

Le Deuff, O. (2017). Le chercheur en humanités digitales : un cas particulier de travailleur du savoir? *Communication & management, 14*(1), 55-69. https://doi.org/10.3917/comma.141.0055

Lemieux, C. (2018). I. Principes.Dans *La sociologie pragmatique* (p. 7-35). La Découverte. https://www.cairn.info/la-sociologie-pragmatique--9782707173355-p-7.htm

Le Moigne, J.-L. (1994). *La théorie du système général : théorie de la modélisation* (4e éd., mise à jour). PUF.

Lemonnier, P. (1983). À propos de Bertrand Gille : la notion de « système technique ». *Homme*, *23*(2), 109-115. https://doi.org/10.3406/hom.1983.368375

Lugan, J.-C. (2009). *La systémique sociale.* Presses universitaires de France.

Massot, M.-L., Tricoche, A., et Mulas, S. (2023, 30 juillet). *Revues en humanités numériques* [Matériels pédagogiques]. Digit_Hum; Digit_Hum. https://digithum.huma-num.fr/ressources/revues/

MENJ (2023a). *Enseigner et apprendre avec la recherche : les groupes thématiques numériques (GTnum)*. éduscol | Ministère de l'Éducation nationale et de la Jeunesse, Direction générale de l'enseignement scolaire. http://tinyurl.com/584wu7ks

MENJ (2023b, janvier). *Stratégie du numérique pour l'éducation 2023-2027*. Ministère de l'Éducation nationale et de la Jeunesse. https://www.education.gouv.fr/strategie-du-numerique-pour-l-education-2023-2027-344263

MENJ (2023c, mai). *Les incubateurs académiques*. éduscol. https://eduscol.education.fr/2669/les-incubateurs-academiques

Merzeau, L. (2013). L'intelligence des traces. *Intellectica. La revue de l'Association pour la Recherche sur les sciences de la Cognition (ARCo)*, *1*(59), p.115-135. https://halshs.archives-ouvertes.fr/halshs-01071211/document

Merzeau, L., et Mulot, H. (2017). Les communs : Levier pour l'enseignement (du) numérique à l'école. *Hermès, La Revue*, *78*, 193-200. http://www.cairn.info/revue-hermes-la-revue-2017-2-p-193.htm

Ministère de l'Enseignement supérieur et de la Recherche (2023). *scanR* | *Moteur de la Recherche et de l'Innovation*. ScanrR. https://scanr.enseignementsup-recherche.gouv.fr/

Moeglin, P. (2010). *Les industries éducatives*. Presses universitaires de France.

Monod-Ansaldi, R., Vincent, C., et Aldon, G. (2019). Objets frontières et brokering dans les négociations en recherche orientée par la conception. *Éducation didactique*, *13*(2), 61-84. https://www.cairn.info/revue-education-et-didactique-2019-2-page-61.htm

Ngram Viewer. (2023, 31 août). *TIC,TICE,informatique,numérique*. Google Books Ngram Viewer. http://tinyurl.com/mspymwra

OCDE-CERI (1971). *La technologie de l'enseignement : conception et mise en œuvre de systèmes d'apprentissage*. OCDE.

OECD (2023). *Shaping Digital Education: Enabling Factors for Quality, Equity and Efficiency*. Organisation for Economic Co-operation and Development. https://www.oecd-ilibrary.org/education/shaping-digital-education_bac4dc9f-en

Pélisset, É. (1985). *Pour une histoire de l'informatique dans l'enseignement français*. Edutice Archives ouvertes. https://edutice.hal.science/file/index/docid/276158/filename/h85ep.htm

Pestre, D. (2003). *Science, argent et politique*. Editions Quae.

Plantard, P. (2014). *Anthropologie des usages du numérique* [thèse, Université de Nantes, France]. https://halshs.archives-ouvertes.fr/tel-01164360/document

Plantard, P. (2023, juin 29). *L'appropriation des technologies numériques par les enseignants : processus, modalités et contextes* (INSPE de Bretagne, Bretagne, France) [Vidéo]. Pod Inspé Bretagne; INSPE de Bretagne. http://tinyurl.com/33455d6b

Rabardel, P. (1995). *Les hommes et les technologies, approche cognitive des instruments contemporains*. Armand Colin. https://hal.archives-ouvertes.fr/hal-01017462

Région académique Guyane (2023). Feuille de route pour le développement des usages numériques. Délégation de Région Académique au Numérique pour l'Éducation. http://tinyurl.com/3tdnm8st

Renaud, L. (2020). Modélisation du processus de la recherche participative. *Communiquer. Revue de communication sociale et publique*, *30*. https://doi.org/10.4000/communiquer.7437

Romero, M., Lepage, A., et Lille, B. (2017). Computational thinking development through creative programming in higher education. *International Journal of Educational Technology in Higher Education*, 14. https://doi.org/10.1186/s41239-017-0080-z

Rouissi, S. (2017). L'apparition du numérique dans les discours officiels sur l'école en France. *Hermès, La Revue*, *78*, 31-40. http://www.cairn.info/revue-hermes-la-revue-2017-2-p-31.htm

Roux, J. (2004). Penser le politique avec Simondon. *Multitudes*, *18*(4), 47-54. https://doi.org/10.3917/mult.018.0047

Simondon, G. (1989). *Du mode d'existence des objets techniques* (Éd. augmentée). Aubier.

Stiegler, B. (dir.). (2014). *Digital studies : Organologie des savoirs et technologies de la connaissance.* Entretiens du nouveau monde industriel. Fyp éditions et IRI, Institut de recherche et d'innovation.

UNESCO (2023). *Global education monitoring report, 2023: Technology in education: a tool on whose terms?* https://doi.org/10.54676/UZQV8501

Vitali-Rosati, M. (2014). *Pour une définition du « numérique »*. Les Presses de l'Université de Montréal. https://papyrus.bib.umontreal.ca/xmlui/handle/1866/13162

Voulgre, E. (2011). *Une approche systémique des TICE dans le système scolaire français : entre finalités prescrites, ressources et usages par les enseignants* [thèses, Université de Rouen]. https://hal.archives-ouvertes.fr/tel-01628569

Voulgre, E. (2012, juillet 2). Une approche systémique des technologies de l'information et de la communication en éducation dans le système scolaire français : entre finalités prescrites, ressources et usages par les enseignants : proposition d'une synthèse. *Adjectif*. https://adjectif.net/spip.php?article157

Wallet, J. (2010). Technologie et gouvernance des systèmes éducatifs. Dans B. Charlier et F. Henri (dir.), *Apprendre avec les technologies*. Presses universitaires de France.